# Influence of the losses on the super-resolution performances of an impedance matched negative index material

**Giuseppe D'Aguanno[†], Nadia Mattiucci and Mark J. Bloemer**
*Charles M. Bowden Research Center, RDECOM, Building 7804, Redstone Arsenal, AL 35898-5000, USA.*

**Abstract**

We discuss by a Poynting vector analysis how the losses of a negative index material (NIM) affect the resolution performances of a Veselago-Pendry lens and we analyze those performances in the framework of the Abbe criterion. Both the limit of high losses and low losses are explored. We find that the impedance matched NIM is able to resolve 30% better than the limit imposed by the Abbe criterion even when the imaginary part of the refractive index (the material losses) exceeds the absolute value of the real part of the refractive index. The NIM is described by a lossy Drude model with equal permittivity and permeability. By increasing the damping parameter of the Drude model we also explore the regime where both permittivity and permeability are positive and point out the conditions under which the metamaterial is still able to super-resolve.

[†] Corresponding Author: giuseppe.daguanno@us.army.mil *or* giuseppe.daguanno@gmail.com; tel. 001-256-8429815 or 001-256-955-6955; fax 001-256-9557216

## 1. Introduction

In the past few years, negative index materials (NIMs), i.e. materials that have simultaneously negative permittivity and magnetic permeability [1], have been the subject of intense theoretical and experimental investigations [2-5]. Several applications have been envisioned for those materials [6-9], although, arguably, the most important is the possibility to use them to construct a "perfect" lens, i.e. a lens that can also focus the evanescent near-field components of an object, as pointed out by Pendry several years ago in his seminal paper [2]. One serious issue that plays a detrimental role toward the achievement of a super-resolving lens is the fact that in currently available meta-materials the absorption is still very high. In order to characterize the quality of a NIM it is helpful to introduce its Figure of Merit (FOM) defined as: $FOM = -\mathrm{Re}(\hat{n})/\mathrm{Im}(\hat{n})$, where $\hat{n}$ is the complex refractive index of the NIM. The path towards the realization of NIMs in the near infrared and visible regime with high FOMs is still long, although remarkable progresses have been made very recently. In 2005 the first NIMs operating in the visible regime were reported [10-11] and shortly after a low-loss silver-based NIM operating at telecommunication wavelengths was theoretically studied [12] and experimentally realized [13] with a FOM~3 at ~1.5μm. Recently, negative index metamaterials operating respectively at 780nm with a FOM~0.5 [14] and at 813nm with a FOM~1 [15] have been realized and tested.

The aim of this work is to twofold: first, to study how the losses and the FOM of the NIM influence its capability to act as a super-resolving lens; second, to assess what is in principle the maximum resolution power achievable in a NIM-based lens. As a prototype and benchmark of a NIM-based device for super-resolution purposes we study the configuration described in Fig.1. In Fig.1 the distance between the object plane and the input surface of the lens $d_1$ and the distance of the image plane from the output surface of the lens $d_2$ are chosen according to the geometrical rule: $d_2=L-d_1$, where L is the thickness of the NIM slab. This is the geometrical rule that assures the optimum image formation for a flat lens with refractive index n=-1, as pointed out in Ref.[2].

In order to avoid any ambiguity we have chosen to study two point-like sources: i.e. we consider two slits in the object plane whose dimension is *~1/500$^{th}$* of the incident wavelength $\lambda$ so that, for all intent and purposes, the limit of point source is reached. The mutual distance *D* between the two points is therefore the only parameter that characterizes the source. The calculations of the diffraction figures on the image plane have been carried out using the technique of the angular spectrum decomposition in conjunction with the transfer matrix technique [16]. More details about the calculation technique can be found in the Appendix. The permittivity and permeability of the NIM are described with a degenerate, lossy Drude model [2]: $\hat{\varepsilon}(\tilde{\omega}) = \hat{\mu}(\tilde{\omega}) = \hat{n}(\omega) = 1 - 1/[\tilde{\omega}(\tilde{\omega} + i\tilde{\gamma})]$, $\tilde{\omega} = \omega / \omega_{ep}$ is the normalized frequency, $\omega_{ep} = 2\pi c/\lambda_{ep}$ is the electric plasma frequency and $\lambda_{ep}$ the corresponding wavelength, $\tilde{\gamma} = \gamma / \omega_{ep}$ is the damping term normalized with respect to the electric plasma frequency. Note that here we are supposing that the electric and magnetic plasma frequency are equal, i.e. $\omega_{ep}=\omega_{em}$. The condition $\hat{\varepsilon} = \hat{\mu} = \hat{n}$ obviously implies that the impedance of the medium $Z = \sqrt{\hat{\mu}/\hat{\varepsilon}}$ is perfectly matched with the impedance of vacuum (Z=1) for any value of the damping term and any value of the refractive index. Note also that, while we are always in a condition of impedance matched, the particular condition n=-1 has the further advantage with respect to the simple condition of impedance matched to reduce the reflection for a wider range of incident angles. Among metamaterials with super-resolution capabilities, impedance matched metamaterials play an important role because their response is independent from the particular polarization of the incident light and therefore the super-resolving capabilities are the same both for TE and TM light polarization. For instance, in Ref.[17] Aydin et al. report an impedance-matched, low loss negative-index metamaterial superlens operating in the microwave regime (3.74GHz) that is capable of resolving subwavelength features of a point source with a 0.13λ resolution.

We fix the operative wavelength at $\lambda = \sqrt{2}\lambda_{ep}$ so that the real part of the refractive index of the NIM is approximately –1 for low values of the damping term. For example, at $\lambda$ ~360nm the length of the NIM slab and the distances between the object plane an the image plane are the same as those reported in Ref.[2] for the super-resolving lens described by Pendry. At the operative wavelength $\lambda = \sqrt{2}\lambda_{ep}$ the expression of the refractive index and of the FOM are the following: $Re(\hat{n}) = \frac{-1+2\tilde{\gamma}^2}{1+2\tilde{\gamma}^2}$, $Im(\hat{n}) = \frac{2\sqrt{2}\tilde{\gamma}}{1+2\tilde{\gamma}^2}$, $FOM = \frac{1-2\tilde{\gamma}^2}{2\sqrt{2}\tilde{\gamma}}$. In Figs.2 we show the FOM of the NIM and its refractive index at the operative wavelength $\lambda = \sqrt{2}\lambda_{ep}$ as function of the damping term, $\tilde{\gamma}$ . The figures clearly identify two regimes: 1) for $\tilde{\gamma} < 1/\sqrt{2}$ we have that $\mathrm{Re}(\hat{n}) < 0$ and FOM>0, i.e. the metamaterial acts as a true NIM; 2) for $\tilde{\gamma} > 1/\sqrt{2}$ we have that $\mathrm{Re}(\hat{n}) > 0$ and FOM<0, i.e. the metamaterial acts as a positive index material (PIM). For $\tilde{\gamma} = 1/\sqrt{2}$ the refractive index is a pure imaginary number: $\hat{\varepsilon} = \hat{\mu} = \hat{n} = i$ and the metamaterial acts as an impedance matched material having the real part of the refractive index equal to zero. For the sake of clarity we have also reported in Fig.2(a) the region where the metamaterial super-resolves, which is one of the main results of our analysis. We will give the details below, suffice here to note that the metamaterial super-resolves even in the region where it behaves as a PIM. Before going into the details of our analysis it is worth to spend some words to recall the classical limit under which two point-like objects cannot be resolved, this limit is sometime also known as the “Abbe criterion” [18]. The Abbe criterion states that the smallest distance between two point-like objects that an optical instrument, such as a classical microscope, can resolve is $\sim 0.6\lambda/(n \sin\theta)$ where n is the refractive index of the immersion medium and $\theta$ the is the half-angle subtended by the object at the lens [18]. The two objects can be considered resolved when the image contrast or visibility ($V$) of the diffraction figure is at least ~50%, where $V(\%)=100(I_{max}-I_{min})/(I_{max}+I_{min})$, $I_{max}$ and $I_{min}$ are respectively the maximum and minimum intensity of the diffraction figure. In principle the Abbe criterion does not prevent sub-wavelength resolution. In our case the immersion medium is air, therefore even assuming an infinitely extended lens so that $\theta\sim\pi/2$ we have that the smallest distance is ~0.6λ. It is also worthwhile to spend few words regarding the properties of Pendry's perfect lens [2]. There are two equally important characteristics that contribute to make Pendry's lens "perfect". The first one is that the lens is flat and it has a refraction index n=-1, therefore, speaking in terms of geometrical optics, it focuses all the rays only in one point avoiding the problem of spherical aberrations which are typical of conventional curved-lens. The second characteristic is that the lens is able to reconstruct or amplify the near field components of the object giving therefore the possibility to beat the Abbe criterion. One of the findings of this work is that actually the possibility to beat the Abbe criterion can be achieved even when the lens is significantly off the ideal condition of n=-1 and therefore its focusing properties are less than ideal because aberrations come into play. On the other hand, the condition of ideal focusing alone without the reconstruction or amplification of the evanescent modes seems not to be sufficient to beat the Abbe criterion. At this regards an outstanding example of a flat lens in the form of a two-dimensional photonic crystal with an effective index n=-1 and almost ideal focusing properties, but negligible reconstruction of the evanescent modes, can be found in Ref.[19]. In the work cited in Ref.[19] sub-wavelength resolution but not super-resolution is achieved, i.e. the lens does not beat the Abbe criterion. Keeping in mind the above considerations, let us now go to analyze our results. The paper is organized as follows: in Section 2 we analyze the regime where $\tilde{\gamma} < 1/\sqrt{2}$, in Section 3 the regime where $\tilde{\gamma} > 1/\sqrt{2}$, in Section 4 we analyze the realistic case of a single layer of silver and finally in Section 5 we go to the conclusions.

**2. Super-resolution for $\tilde{\gamma} \leq 1/\sqrt{2}$ (FOM≥0).**

As we have already mentioned in the introduction, in the case at hand the metamaterial acts as a true NIM with $\mathrm{Re}(\hat{n}) < 0$ and FOM>0. First we analyze the case when the square modulus of the fields is considered, electric field for TE polarization or magnetic field for TM polarization. In Fig.3 we show the diffraction figures and the minimum distance ($D_{Min}$) which the NIM lens is able to resolve (according to a field analysis) for a FOM respectively of ~3500 (Fig.3(a)), ~350 (Fig.3(b)), ~35 (Fig.3(c)), and ~3.5 (Fig.3(d)). In Fig.4 we show $1/D_{Min}$ as function of the FOM for values of the FOM ranging from $10^{-4}$ to $10^{4}$. The figure suggests that the scaling law of the minimum resolved distance as function of the FOM is as follows:

$$D_{Min} \cong \begin{cases} \dfrac{\lambda}{2\log(FOM)} \cong \dfrac{\lambda}{2\left|\log(\mathrm{Im}(\hat{n}))\right|} & for \;\; FOM >> 1 \,,\; \mathrm{Im}(\hat{n}) << 1 \\ 0.4\lambda & for \quad 0 \leq FOM << 1 \,,\; \mathrm{Im}(\hat{n}) \cong 1 \end{cases} \quad . \qquad (1)$$

The losses adversely affect the super-resolution process because they lower both the transmission resonance of the evanescent modes and the transmission of the propagative modes. It is important to point out that an analysis of the influence of the losses on the super-resolution capabilities of a NIM has also been studied in Ref. [20] where a logarithmic dependence of the minimum resolved distance has been analytically calculated in the limit of low losses ($\mathrm{Im}(\hat{n}) << 1$).Our analysis differs from that one for two fundamental aspects: first, we analyze also the cases of high losses, and, second, we asses our results based on a power analysis (time-averaged Poynting vector) rather than on the electric or magnetic field separately. At this regard, we would like to point out, using the words of Born and Wolf [21], that *"in optics the (averaged) Poynting vector is the chief quantity of interest"*. As we have already mentioned in the Introduction, currently available metamaterials are still characterized by high losses and low FOMs [11-15] and therefore a sound theoretical analysis, based on the Poynting vector, of the limit of high losses seems at order. In fact, as it will become clear later, our analysis shows that absorption does not always play a detrimental role for super-resolution purposes.

In Fig.5 we show the diffraction figures and the minimum resolved distance when the z-component of the Poynting vector ($S_z$) is used for a FOM respectively of ~3500 (Fig.5(a)), ~350 (Fig.5(b)), ~35 (Fig.5(c)), and ~3.5 (Fig.5(d)). The fact that $S_z$ is negative on the lateral parts of the diffraction figures is due to the interference of propagating and evanescent modes which can cause circulation in the Poynting vector, as previously noted [22]. In Fig.6 we show the scaling law for $D_{Min}$ which is the following:

$$D_{Min} \cong \begin{cases} \dfrac{\lambda}{1.8\log(FOM)} \cong \dfrac{\lambda}{1.8\left|\log(\mathrm{Im}(\hat{n}))\right|} & for \quad FOM >> 1 \,,\ \mathrm{Im}(\hat{n}) << 1 \\ 0.4\lambda & for \quad 0 \le FOM << 1 \,,\ \mathrm{Im}(\hat{n}) \cong 1 \end{cases} . \qquad (2)$$

For $FOM<<1$ $D_{min}$ still scales as $0.4\lambda$, for high values of the FOM the scaling law is slightly different with respect to the previous case(~10% less in resolving power).

Although in this case the field analysis and the Poynting vector analysis give results qualitatively similar, it should be underlined that this is not always the case and one should always resorts to the Poynting vector in order to asses the super-resolution performances of a metamaterial. We will provide later an example where the field analysis and the Poynting vector analysis give different results. For the time being let us comment Fig.6. For $0 \le FOM<<1$ there is a characteristic plateau at $0.4\lambda$ in the super-resolution capability of the lens which still represents a remarkable improvement of *~30%* over the limit of $0.6\lambda$ imposed by the Abbe criterion. The plateau is reached for values of the damping coefficient $\tilde{\gamma}$ close to $1/\sqrt{2}$. It is interesting that even in the extreme case of $\tilde{\gamma} = 1/\sqrt{2}$ where FOM=0 and $\hat{\varepsilon} = \hat{\mu} = \hat{n} = i$ (i.e. the metamaterial becomes an impedance matched material having a zero real part of the refraction index), the lens still resolves at $0.4\lambda$, i.e. beating the Abbe criterion. In this sense the Figure of Merit not always is a good indicator of the super-resolution capabilities of a NIM. In Fig.7(a) we show the transmittance $|t|^2$ of the slab as function of the transverse wavevector ($k_x$) normalized to the vacuum wavevector ($k_0$) for the case $\tilde{\gamma} = 1/\sqrt{2}$ and in Fig.7(b) we show a 3-D topographic plot of the transmittance vs. $k_x/k_0$ and $\tilde{\gamma}$. As regards Fig.7(a), the transmission resonance at $k_x/k_0>1$, which is the transmission resonance for the evanescent modes, is an indication that the evanescent modes are somewhat "amplified" and they contribute to the image formation with subwavelength details. Of course, amplification of the evanescent modes is just one of the two key elements to obtain super-resolution, the second key element is the suppression of the diffraction of the propagation modes. In this case, differently from the case $\hat{\varepsilon} = \hat{\mu} = \hat{n} = -1$, the rays no longer form a focus inside the lens but they go parallel like in a perfect collimator as shown in Figs.8 and, although the suppression of the diffraction is achieved only in the lens, this is sufficient to obtain image formation. Figs.8(a) and 8(b) exemplify the two limit regimes, i.e. the regime of ideal focusing and the regime of collimation. Of course, by increasing the value of the damping term the lens passes gradually from the ideal focusing to a regime of less than ideal focusing with aberrations coming into play and then it finally approaches the regime of collimation. The reader can easily convince himself that an impedance matched metamaterial having $\mathrm{Re}(\hat{n}) = 0$ acts as a perfect collimator by a simple plane wave analysis. The refraction angle of the Poynting vector ($\vartheta_{\vec{S}}$) for a plane, monochromatic wave incident on our impedance matched metamaterial (i.e. $\hat{\varepsilon} = \hat{\mu} = \hat{n}$) from vacuum at an angle $\vartheta_0$ with respect to the normal at the interface is given by the following formula:

$$\tan(\vartheta_{\vec{S}}) = \frac{\mathrm{Re}(\hat{n})\sin\vartheta_0}{\left([\mathrm{Re}(\hat{n})]^2 + [\mathrm{Im}(\hat{n})]^2\right)\mathrm{Re}\left(\sqrt{1-\frac{\sin^2\vartheta_0}{\hat{n}^2}}\right)} \quad . \qquad (3)$$

From Eq.(3) it follows that $\vartheta_{\vec{S}} = 0$ for any incident angle $\vartheta_0$ whenever $\mathrm{Re}(\hat{n}) = 0$. It must be noted that some properties of impedance matched metamaterials having a zero index of refraction have also been studied in Ref.[23]. In particular, in Ref.[23] the property of those materials to convert wave fronts with small curvature into output beams with large curvature (planar) wave fronts was numerically investigated through finite-difference time domain simulations. A single slab of silver at the wavelength of 324nm has practically a purely imaginary permittivity $\hat{\varepsilon} \cong 0.74i$ [24] and therefore it will behave for TM light polarization similarly to the case of the NIM with a FOM=0 that we have just described. We have in fact calculated the super-resolution of a layer of Ag at the operative wavelength of λ=324nm for the same geometry described in Fig.1 and find that the minimum resolved distance is approximately 0.4λ, i.e. the same as for the case described above.

### 3. Super-resolution for $\tilde{\gamma} > 1/\sqrt{2}$ (FOM<0).

In this case the metamaterial starts to behave as a positive index material with $\mathrm{Re}(\hat{\varepsilon}) = \mathrm{Re}(\hat{\mu}) = \mathrm{Re}(\hat{n}) > 0$ and FOM<0, moreover the imaginary part of the refractive index starts to decrease (see Figs.2). We have varied the damping parameter $\tilde{\gamma}$ in the range $[1/\sqrt{2}, 10]$ that corresponds to varying the FOM approximately in the range [0, -7] and varying $\mathrm{Re}(\hat{\varepsilon}) = \mathrm{Re}(\hat{\mu}) = \mathrm{Re}(\hat{n})$ approximately in the range [0,1] (see Figs.2). In Fig.9 we show *1/Dmin* vs. the absolute value of the *FOM*. Quite surprisingly the figure shows that the metamaterial super-resolves until the FOM reaches the value of approximately -2.37 which gives the minimum resolved distance at ~0.6λ that is the limit of the Abbe criterion. For a FOM=-2.37 the material parameters are respectively: $\tilde{\gamma} \cong 3.5$ and $\hat{\varepsilon} = \hat{\mu} = \hat{n} \cong 0.92 + i0.38$. This is an interesting result because it tells us that even a metamaterial with the real part of the permittivity and permeability close to 1 can in principle super-resolve. The implications for the practical design of a metamaterial are important, in fact basically this means that *negative values of either the permittivity or the permeability (or both of them) are not necessary for super-resolution purposes.* For $\tilde{\gamma} > 3.5$ the transmittance of the evanescent modes is even greater than that for $\tilde{\gamma} \cong 1/\sqrt{2}$ (see Figs.7). In fact for $\tilde{\gamma} > 3.5$ the super-resolution is lost because of the loss of the canalization regime [25], in other words the metamaterial no longer super-resolves because it can no longer compensate for the diffraction of the propagative modes. From Eq.(3) we can infer that the refracted rays go parallel inside the metamaterial for any incident angle, as described in Fig.8(b), whenever one the following conditions is met: a) $\left|\mathrm{Re}(\hat{n})\right| << \mathrm{Im}(\hat{n})$, b) $\left|\mathrm{Re}(\hat{n})\right| >> \mathrm{Im}(\hat{n})$ and $\left|\mathrm{Re}(\hat{n})\right| >> 1$ . The condition $\tilde{\gamma} = 1/\sqrt{2}$ represents the case of perfect collimation because the real part of the refractive index is exactly zero. Increasing the value of the damping over $1/\sqrt{2}$ causes the real part of the refractive index to increase and its imaginary part to decrease (see Figs.2), therefore the effect of the collimation will be gradually lost. In this sense, in the case at hand, the decreasing of the absorption plays two antithetical roles: on one side it improves the transmission of the evanescent modes, but, on the other side, it causes the loss of collimation of the propagative modes. The super-resolution is lost when the beneficial effect of the better transmission of the evanescent modes is not anymore able to compensate the detrimental effect of the loss of collimation. If anything, what our discussion shows is that the imaginary part of the refractive index does not always play a detrimental role, but in some circumstances its role can be even beneficial as regards super-resolution. This finding may be of some help in the practical design of metamaterials. Given the fact that consistent losses in currently available metamaterials seem unavoidable, one may take advantage of those losses exploiting those particular regimes. At this regard, as final example, in the next Section we would like to analyze the realistic case of a single layer of silver.

**4. Single layer of silver**

First of all let us study the spectral region where surface plasmon polariton modes (SPPs) exist for an air/silver interface. The complex dispersion of the SPPs at the air-silver interface is given by [26]:

$$\hat{k}_{SPP} = k_0 \left( \sqrt{\frac{\hat{\varepsilon}}{1+\hat{\varepsilon}}} \right) \quad , \qquad (4)$$

where $\hat{\varepsilon}$ is the complex permittivity of silver and $k_0$ the vacuum wavevector. SPPs are in general guided modes that may exist at the interface dielectric/metal, these modes are also nonradiative in nature, i.e. $\mathrm{Re}(\hat{k}_{SPP}) > k_0$ in the case air/metal. In Fig.10 we show the effective index defined as: $\hat{n}_{SPP} = \hat{k}_{SPP} / k_0$. The spectral region where SPPs are found is defined by the condition $\mathrm{Re}(\hat{n}_{SPP}) > 1$. The values of the permittivity of silver are taken from experimental data [24]. Conventional wisdom would say that a silver layer super-resolves only in a certain region around 337nm where $\mathrm{Re}(\hat{\varepsilon}) \cong -1$ and where SPPs can be excited. The condition $\mathrm{Re}(\hat{\varepsilon}) \cong -1$ assures the compensation of the diffraction through the focusing mechanism described in Fig.8(a), while the excitation of a SPP, given its nonradiative nature, ensures that the structure supports evanescent modes. We now show that a single layer of silver can super-resolve for TM light polarization with a minimum resolved distance of *~λ/3* (*~50%* better than the limit of 0.6λ imposed by the Abbe criterion) for an operative wavelength of *λ=295nm*, i.e. well beyond its plasma frequency (*$\lambda_{ep,Ag}$=328nm (3.78eV)*) where SPPs cannot be excited and the real part of the permittivity is positive. The geometry is described in Fig.11(a). In this case, differently from the geometry of Fig.1, the two slits are located at the entrance of the lens (but in free space) and the image is placed at the end face of the lens. In this way we have the canalization mechanism taking place inside the lens and at the same time, by having the image and the object plane attached respectively at the output and input surface of the lens, we avoid diffraction outside the lens. The distance of the object plane from the image plane is the same fraction of the operative wavelength as in Fig.1. The refractive index of silver at 295nm is $\hat{n} \cong 1.52 + i1.08$ [24] and the permittivity is $\hat{\varepsilon} \cong 1.13 + i3.3$. In Fig. 11(b) we show that the two point sources located in the object plane with a mutual distance of *$D=\lambda/3$* are resolved in the image plane with a visibility *$V\sim 52\%$*, which unambiguously proves super-resolution. It should be therefore clear that the excitation of plasmonic modes is not necessary for super-resolution purposes. Without the Ag layer in place the visibility would be *$V\sim 5\%$* (i.e. one order of magnitude less). The minimum resolved distance without the Ag layer is *$D_{min}\sim 0.8\lambda$*. Again, super-resolution for TM polarization is possible because: a) there is a resonance in the transmittance of the evanescent modes and b) the propagative modes are canalized inside the layer. Let us first analyze the canalization process in this case. A simple analysis can show that the refraction angle of the Poynting vector ($\vartheta_{\vec{S}}$) for a plane, monochromatic wave incident on generic material from vacuum at an angle $\vartheta_0$ with respect to the normal at the interface is given by the following formulas:

$$\tan(\vartheta_{\vec{S}}) = \frac{\mathrm{Re}(\hat{\varepsilon})\sin\vartheta_0}{|\hat{\varepsilon}|^2\,\mathrm{Re}\left(\dfrac{\hat{n}}{\hat{\varepsilon}}\sqrt{1-\dfrac{\sin^2\vartheta_0}{\hat{n}^2}}\right)}\ , \quad \text{(TM polarization)} \quad , \qquad (5.1)$$

$$\tan(\vartheta_{\vec{S}}) = \frac{\mathrm{Re}(\hat{\mu})\sin\vartheta_0}{|\hat{\mu}|^2\,\mathrm{Re}\left(\frac{\hat{n}}{\hat{\mu}}\sqrt{1-\frac{\sin^2\vartheta_0}{\hat{n}^2}}\right)} \text{ , (TE polarization) .} \qquad (5.2)$$

In Fig.12 we show the refracted angle of the Poynting vector as function of the incident angle for the case of Ag at 295nm. It is evident that the canalization process is taking place only for TM polarization of the light, although not in an ideal way. In Fig.13 we show the transmittance vs. $k_x/k_0$. In this case the transmission of the evanescent modes is effective just for TM polarization. The fact that a material with a real part of the refractive index greater than 1 may support evanescent modes should not be surprising after all. Amplification of evanescent waves by two dielectric planar waveguides has been, for example, analytically demonstrated in Ref. [27]. Although the conditions and the geometry studied in Ref.[27] are different from ours, nevertheless the physical reasons underlying the supporting of evanescent waves are essentially the same, i.e. in our case the single layer of Ag at 295nm acts like it were a dielectric waveguide with strong losses, and therefore it is able to couple part of the evanescent modes which become quasi-guided modes inside the layer. In order to be as clear as possible regarding the physical origin of the amplification of the evanescent modes in a slab with n>1, we would like to remark that any time guided modes are excited in the transverse direction (x-axis)-be either plasmonic modes or just conventional guided modes-the transmittance $T(k_x)$ along the z-axis will show resonances (amplification) in the evanescent part of the spectrum exactly at the $k_x$'s where those guided modes are excited. In this sense there is a unifying concept that lays behind the amplification of the evanescent modes in single layer of metal with $\mathrm{Re}(\hat{\varepsilon}) = -1$ for TM polarization and in a simple slab of dielectric material with $\mathrm{Re}(\hat{n}) > 1$: in both cases amplification of the evanescent modes happens when guided modes are excited in the transverse direction, in the first case they are SPP mode, while in the latter case they are conventional guided modes. Of course, in the example we have proposed, the price to pay in order to obtain a good degree of super-resolution ($D_{min}=\lambda/3$) in a region where plasmonic modes are not involved is that the transparency of the lens is not high, only *~5%* of the incident power is transmitted to the other side. We have also calculated the super-resolution with the Ag layer for the same geometry described in Fig.1, i.e. for a distance of $\lambda/18$ between the object (image) plane and the beginning (end) of the lens and a thickness of the Ag layer of $\lambda/9$. Although the TM transmittance for both propagative and evanescent modes greatly improves with respect to the previous case because half of the thickness of Ag is now used, nevertheless the visibility of the image drops to *V~15%* because the canalization process is now less effective due to the fact that it is taking place in just half of the distance between the image plane and the object plane. Finally, we would like to come back to the issue of the field analysis vs. Poynting vector analysis. In Fig.14 we show the square modulus of magnetic field ($|H|^2$ ) and $S_z$ for the case of Fig.11. It is clear that the two quantities differ significantly each other as regards their shape. A field analysis in this case would be inappropriate in order to asses the super-resolution performances of the object.

**5. Conclusions**

In conclusion, we have discussed the influence of the losses of an impedance matched NIM on its super-resolving capability. The NIM has been described by a lossy Drude model with equal permittivity and permeability. The analysis has been performed using the Poynting vector as the chief quantity of interest instead of the electric or magnetic fields separately. We have found that in the limit of low losses $(\mathrm{Im}(\hat{n}) << 1)$ the minimum resolved distance $D_{Min}$ scales as $D_{Min} \cong \lambda / (1.8|\log(\mathrm{Im}(\hat{n}))|)$, while in the limit of high losses $(\mathrm{Im}(\hat{n}) \sim 1)$ $D_{Min} \cong 0.4\lambda$. By varying the damping parameter in the Drude model we have also explored the case where both the permittivity and permeability are positive and we have found that the metamaterial still continues to super-resolve until the real part of the refractive index becomes close to 1. We have discussed the physical mechanisms behind the super-resolution when the permittivity and permeability are both positive. We have also provided an example of super-resolution at ~$\lambda$/3 from a single layer of Ag well beyond its plasma frequency where plasmonic modes are not excited and the real part of the permittivity is ~1. It is also worthwhile noting that in all the cases we have analyzed the super-resolution is lost very rapidly once the image plane is placed one or two wavelengths away from the end of the lens.

We hope that our results may be of some guidance for the future design of metamaterials for super-resolution purposes.

**Acknowledgments**

Giuseppe D'Aguanno and Nadia Mattiucci thank the NRC for financial support. Giuseppe D'Aguanno's e-mail addresses are: giuseppe.daguanno@us.army.mil or giuseppe.daguanno@gmail.com

**Appendix A**

In this Appendix we give full details on the calculation method used. We consider non-dimensional units, i.e. we take $\varepsilon_0=\mu_0=c=1$ ($\varepsilon_0$ and $\mu_0$ are respectively the permittivity and permeability of vacuum and $c$ is the speed of light in vacuum). Let us start by discussing the calculation method for TM polarization of the incident light. The calculations of the diffraction figures on the image plane have been carried out using the technique of the angular spectrum decomposition [28] in conjunction with the transfer matrix technique [29]. Referring to Fig.1, a plane, monochromatic, TM polarized wave of unitary amplitude and with wave-vector $k_0=2\pi/\lambda$ is incident on the object plane which is at a distance $d_1$ from the input surface of the NIM slab of length $L$. The input surface of the NIM slab is located at z=0 along the z-axis, the object plane is located at $z=-d_1$and the image plane is located at $z=L+d_2$ . In our reference frame, the magnetic field diffracted from the object plane is expressed as:

$$\tilde{\vec{H}}(x,z,t)=(1/2)\left[\vec{H}(x,z)\exp(-i\omega t)+c.c\right] , \qquad \text{(A.1)}$$

where $\vec{H}(x,z)=H(x,z)\hat{y}$ is the complex, stationary vector field, $\hat{y}$ is the unit vector of the y-axis, c.c. stands for complex conjugate, and the reference system (x,y,z) forms a right-handed tern. The complex amplitude of the magnetic field $H(x,z)$ diffracted in the semi-space beginning at the output of the lens, i.e. at $z \geq L$, is expressed by the following integral [28]:

$$H(x,z\geq L)=\int_{-\infty}^{+\infty} A(k_x)t_{TM}(k_x)\exp[i(k_x x+\sqrt{k_0^2-k_x^2}(d_1+z-L))]dk_x \quad . \qquad \text{(A.2)}$$

Here $k_x$ is a real quantity and it represents physically the wave-vector of the x-axis, $t_{TM}(k_x)$ is the complex transmission function of the NIM slab for TM polarization, transmission function which has been calculated using a matrix transfer technique [29]. Because we are dealing with a single slab placed in vacuum, the transmission function can be explicitly expressed in following form:

$$t_{TM}(k_x) = \frac{2}{2\cos\left(\hat{n}\sqrt{k_0^2 - \frac{k_x^2}{\hat{n}^2}}L\right) - i\left(\frac{\hat{n}\sqrt{k_0^2 - \frac{k_x^2}{\hat{n}^2}}}{\hat{\varepsilon}\sqrt{k_0^2 - k_x^2}} + \frac{\hat{\varepsilon}\sqrt{k_0^2 - k_x^2}}{\hat{n}\sqrt{k_0^2 - \frac{k_x^2}{\hat{n}^2}}}\right)\sin\left(\hat{n}\sqrt{k_0^2 - \frac{k_x^2}{\hat{n}^2}}L\right)} \quad . \qquad \text{(A.3)}$$

The magnetic field at the image plane can be calculated from Eq.(A.2) by putting z=L+$d_2$. Note that the integral (A.2) extends over both the propagative ($|k_x| \le k_0$) and evanescent modes ($|k_x| > k_0$). $A(k_x)$ is the Fourier spectrum of the magnetic field on the object plane. In particular, in our case, $A(k_x)$ is nothing else than the Fourier transform (FT) of the transmission function of the screen located at z=-$d_1$:

$$A(k_x) = FT(t_{screen}(z = -d_1, x)) \quad , \qquad \text{(A.4)}$$

where the transmission function of the screen is defined as:

$$t_{screen}(z = -d_1, x) = \begin{cases} 0 & -\infty < x < -D/2 - a_1 \\ 1 & -D/2 - a_1 \le x \le -D/2 \\ 0 & -D/2 < x < D/2 \\ 1 & D/2 \le x \le D/2 + a_2 \\ 0 & D/2 + a_2 < x < \infty \end{cases} \quad . \qquad \text{(A.5)}$$

In practice, the transmission of the screen describes two slits respectively of width $a_1$ and $a_2$ located at a mutual distance (center to center) of D+($a_1$+$a_2$)/2. Basically Eq.(A.5) is the sum of two rectangular (“rect”) functions. The Fourier transform of Eq.(A.5) can, of course, be performed analytically in terms of a linear superposition of sine cardinal (“sinc”) functions. In our case we have taken the width of the two slits so that $a_1=a_2=\lambda/500$ where $\lambda$ is the wavelength of the incident radiation; in this way, for all intents and purposes, the two slits can be considered as point-like sources. Once calculated the complex amplitude of the magnetic field by mean of the integral expressed in (A.2), we can calculate the complex amplitude of the electric field by applying the “curl” operator to the magnetic field as follows:

$$\nabla \times \vec{H} = -i\omega\vec{E} \quad . \tag{A.6}$$

Eq.(A.6) is the differential form of the generalized Ampere law in vacuum for time-harmonic fields in their complex representation [18]. Once calculated the complex amplitude of the electric field, the Poynting vector can be calculated by the well known formula [29]:

$$\vec{S} = (1/2)\,\mathrm{Re}[\vec{E} \times \vec{H}^*] \quad . \tag{A.7}$$

The calculations for a TE polarized wave follow the same procedure outlined above except that the integral (A.2) is calculated for the electric field, the transmission of the NIM slab is for TE polarization and the calculation of the magnetic field from the electric field is done by using the differential form of Faraday's law of induction. Note also that the transmission for TE polarization can be obtained using Eq.(A.3) with the formal substitution $\varepsilon \to \mu$. In the particular case of an impedance matched medium, i.e. $\hat{\varepsilon} = \hat{\mu} = \hat{n}$, the transmission function $t(k_x)$ for TE and for TM polarization is the same.

We would like to underline that the angular spectrum representation applied to layered structures is a semi-analytical technique which relies on the calculation of a simple one-dimensional scattering integral containing well-behaved functions. In problems of diffraction in the near field (especially when plasmon modes are involved) this technique avoids spurious effects which might arise using full numerical simulations as pointed out in Refs.[30-31].

**References and Notes**

[1] V.G. Veselago, "The Electrodynamics of substances with simultaneously negative values of ε and μ", *Sov. Phys. USPEKHI* **10**, 509-514 (1968).

[2] J.B. Pendry, "Negative Refraction Makes a Perfect Lens", *Phys. Rev. Lett*. **85,** 3966-3969 (2000) and references therein

[3] R.A. Shelby, D.R. Smith, and S. Schultz, "Experimental Verification of a Negative Index of Refraction", *Science* ***292***, 77-79 (2001).

[4] C.G. Parazzoli, R. B. Greegor, K. Li, K. E. C. Koltenbah, M. Tanielian, **"**Experimental Verification and Simulation of Negative Index of Refraction Using Snell's Law", *Phys. Rev. Lett.* **90,** 107401-1-4 (2003)

[5] S. Linden, C. Enkrich, M. Wegener, J. Zhou, T. Koschny, C.M. Soukoulis, "Magnetic Response of Metamaterials at 100 Terahertz", *Science* **306,** 1351-1353 (2004).

[6] G.D'Aguanno, N. Mattiucci, M. Scalora, M.J. Bloemer, "Bright and Dark Gap Solitons in a Negative Index Fabry-Perot Etalon" *Phys. Rev. Lett.* **93,** 213902-1-4 (2004).

[7] G. D'Aguanno, N. Mattiucci, M. Scalora, M.J. Bloemer, "TE and TM guided modes in an air waveguide with a negative-index-material cladding" *Phys. Rev. E* **71,** 046603-1-7 (2005).

[8] G. D'Aguanno, N. Akozbek, N. Mattiucci, M. Scalora, M.J. Bloemer, A.M. Zheltikov, *Opt. Lett.* **30,** "Dispersion-free pulse propagation in a negative-index material", 1998-2000 (2005)

[9]M. Bloemer, G. D'Aguanno, M. Scalora, N. Mattiucci, "Broadband omnidirectional reflection from negative index materials", *Appl. Phys. Lett.* **87,** 261921-1-3 (2005)

[10] S. Zhang, W. Fan, N. C. Panoiu, K. J. Malloy, R. M.Osgood, and S. R. J. Brueck, "Experimental Demonstration of Near-Infrared Negative-Index Metamaterials", *Phys. Rev. Lett.* **95**, 137404-1-4 (2005).

[11] V. M. Shalaev, W. Cai, U. K. Chettiar, H. Yuan, A. K. Sarychev, V. P. Drachev, and A. V. Kildishev, "Negative index of refraction in optical metamaterials", *Opt. Lett*. **30**, 3356-3358 (2005).

[12] S. Zhang, W. Fan, K. J. Malloy, S. R. J. Brueck, N. C. Panoiu, and R. M. Osgood, "Near-infrared double negative metamaterials", *Opt. Expr.* **13**, 4922-4930 (2005).

[13] G. Dolling, C. Enkrich, M. Wegener, C.M. Soukoulis, S. Linden,"Low-loss negative-index metamaterial at telecommunication wavelengths", *Opt. Lett*. **31**, 1800-1802 (2006).

[14] G. Dolling, M. Wegener, C.M. Soukoulis, S. Linden, "Negative-index metamaterial at 780 nm wavelength", *Opt. Lett.* **32,** 53-55 (2007)

[15] U. K. Chettiar, A. V. Kildishev, H.-K. Yuan, W. Cai, S. Xiao, V. P. Drachev, and V. M. Shalaev, "Dual-band negative index metamaterial: double negative at 813 nm and single negative at 772 nm", *Opt. Lett.* **32,** 1671 (2007).

[16] M.J. Bloemer, G. D'Aguanno, N. Mattiucci, M. Scalora and N. Akozbek, "Broadband super-resolving lens with high transparency for propagating and evanescent waves in the visible range", *Appl. Phys. Lett.* **90,** 174113-1-3 (2007)

[17] Koray Aydin, Irfan Bulu, and Ekmel Ozbay, "Subwavelength resolution with a negative index metamaterial superlens", *Appl. Phys. Lett.* **90,** 254102-1-3 (2007)

[18] Max Born and Emil Wolf, "Principles of Optics", 7$^{th}$ (expanded) edition, Cambridge University Press, 1999

[19] X.Wang, Z.F. Ren and K. Kempa, "Unrestricted superlensing in a triangular two-dimensional photonic crystal", *Opt. Expr.* **12**, 2919-2924 (2004)

[20] R. Merlin, "Analytical solution of the almost-perfect-lens problem", *Appl. Phys. Lett.* **84**, 1290-1293 (2004) and references therein

[21] Max Born and Emil Wolf, "Principles of Optics", 7th (expanded) edition, Cambridge University Press, 1999, page 10

[22] M-C Yang and K.J. Webb, "Poynting vector analysis of a superlens", *Opt. Lett.* **30**, 2382-2384 (2005)

[23] R.W. Ziolkowski, "Propagation in and scattering from a matched metamaterial having a zero index of refraction", *Phys. Rev. E* **70,** 046608-1-12 (2004)

[24] *Handbood of Optical Constants of Solids,* edited by E.D. Palik, (Academic, New York, 1985).

[25] The term "canalization" has been first used in: P.A. Belov, C.R. Simovski, P. Ikonen, "Canalization of subwavelength images by electromagnetic crystals", *Phys. Rev. B* **71**, 193105-1-4 (2005)

[26] H. Raether, "Surface Plasmons", Springer-Verlag, Berlin (1988)

[27] M. Tsang and D. Psaltis, "Reflectionless evanescent-wave amplification by two dielectric planar waveguides", *Opt. Lett.* **31**, 2741-2743 (2006)

[28] *Optical Coherence and Quantum Optics*, L. Mandel and E. Wolf, (Cambridge University Press, 1995).

[29] *Optical Waves in Crystals,* A. Yariv, P. Yeh, (John Wiley & Sons, New York, 1984).

[30] Y. Zhao, P. Belov, and Y. Hao, "Accurate modeling of the optical properties of left-handed media using a finite-difference time-domain method", *Phys. Rev E* **75,** 037602 (2007)

[31] A. A. Sukhorukov, I.V. Shadrivov, and Yu. S. Kivshar, "Wave scattering by metamaterial wedges and interfaces", Int. Jour. Num. Model. **19,** 105 (2006)

**Figure Captions**

**Fig.1:** A plane, monochromatic wave at the operative wavelength $\lambda$ is incident on a screen of negligible thickness with two very small apertures or slits (P1 and P2) that act as point-like sources (dimension of *~1/500^th^* of the incident wavelength) whose mutual distance is $D$. A slab of NIM L=$\lambda/9$ in length, placed at a distance $d_1=\lambda/18$ from the object plane, captures the light diffracted from the two slits and focus it on the image plane placed at a distance $d_2=\lambda/18$ from the end of the NIM slab. The image plane is chosen following the geometrical rule for the image formation: $d_2$=L-$d_1$.

**Fig.2:** (a) Real (solid line) and imaginary part (long-dashed line) of the refractive index vs. the damping term ($\tilde{\gamma}$) for a wavelength $\lambda = \sqrt{2}\lambda_{ep}$. The horizontal line with double arrows indicates the region where the metamaterial super-resolves, i.e. when it resolves two point sources whose mutual distance is less than $0.6\lambda$ with a visibility $V\sim 50\%$. The super-resolving region extends from $\tilde{\gamma} = 0^+$ to $\tilde{\gamma} \cong 3.5$ which correspond respectively to a refractive index of $\hat{n} = -1 + i0^+$ and $\hat{n} \cong 0.92 + i0.38$. (b) Figure of Merit (FOM) vs. the damping term ($\tilde{\gamma}$). In both figures the short-dashed vertical line indicates the position of $\tilde{\gamma} = 1/\sqrt{2}$, which corresponds to the transition of the metamaterial from a true NIM to a positive index material (PIM).

**Fig.3:** Diffraction figures from the two point-like sources on the image plane respectively for a *FOM~3500* and $D_{Min}=\lambda/16.5$ 4(a), *FOM~350* and $D_{Min}=\lambda/12$ *4*(b), FOM~35 and $D_{Min}=\lambda/8$ *4*(c), *FOM~3.5* and $D_{Min}=\lambda/4.5$ 4(d). The image contrast ($V$) is approximately *50%* in all cases. The square modulus of the field is referred to the electric field for TE polarization or to the magnetic field for TM polarization. $D_{Min}$ is the minimum resolved distance which corresponds to a visibility *V~50%.* In all the figures the maximum value of the field at the image plane has been normalized to 1.

**Fig.4:** $1/D_{Min}$ (solid circles) vs. *FOM* according to a field analysis. For *FOM>>1* the minimum resolved distance ($D_{Min}$) scales as $\lambda/(2log(FOM))$, while for *F<<1* the minimum resolved distance scales as $0.4\lambda$. The dashed, horizontal line represents the limit imposed by the Abbe criterion.

**Fig.5** Diffraction figures calculated through the Poynting vector for a *FOM~3500* and $D_{Min}=\lambda/15$ 4(a), *FOM~350* and $D_{Min}=\lambda/11$ 4(b), *FOM~35* and $D_{Min}=\lambda/7.5$ 4(c), FOM~3.5 and $D_{Min}=\lambda/4.5$ 4(d). In all the figures the maximum of the z-component of the Poynting vector ($S_z$) has been normalized to *1*.

**Fig.6:** $1/D_{Min}$ (solid circles) vs. *FOM* when the Poynting vector is considered. For *FOM>>1* the minimum resolved distance ($D_{Min}$) scales as $\lambda/(1.8log(FOM))$, while for *F<<1* the minimum resolved distance scales as $0.4\lambda$. The dashed, horizontal line represents the limit imposed by the Abbe criterion.

**Fig.7:** (a) Transmittance $|t|^2$ of the NIM vs. $k_x/k_0$ for $\tilde{\gamma}=1/\sqrt{2}$. (b) 3-D topographic plot of the transmittance vs. $k_x/k_0$ and $\tilde{\gamma}$. The vertical solid line indicates the position of $\tilde{\gamma}=1/\sqrt{2}$ where $\mathrm{Re}(\hat{n})=0$.

**Fig.8:** Schematic picture of the rays inside a slab with: (a) $\hat{\varepsilon}=\hat{\mu}=\hat{n}=-1$ and (b) $\hat{\varepsilon}=\hat{\mu}=\hat{n}=i$.

**Fig.9:**$1/D_{Min}$ (solid circles) vs. $|FOM|$ for *FOM<0.* The dashed, horizontal line represents the limit imposed by the Abbe criterion. Also in the figure is indicated the refractive index of the metamaterial at some particular points.

**Fig.10:** Effective index $\mathrm{Re}(\hat{n}_{eff})$ and extinction coefficient $\mathrm{Im}(\hat{n}_{eff})$ vs. wavelength for SPPs at the interface air/silver. The shaded area indicates the region where SPPs exsist. In the figure are also indicated the wavelength of 337nm where $\mathrm{Re}(\varepsilon_{Ag}) \cong -1$ and the wavelength of 295nm where $\mathrm{Re}(\varepsilon_{Ag}) \cong 1.13$.

**Fig.11:** (a) Geometry used for super-resolution from a single layer of silver at λ=295nm. (b) Image of the two point-like sources for TM polarization. The visibility is approximately 52%. The vertical dashed lines indicate the position of the point-like sources in the object plane. $S_z$ is normalized to 1.

**Fig.12:** Refraction angle of the Poynting vector vs. incident angle at the interface air/Ag for a wavelength of 295nm. The dashed line is for TE polarization and the solid line for TM polarization.

**Fig.13:** Transmittance $|t|^2$ of the silver layer at 295nm vs. $k_x/k_0$.

**Fig.14:** Comparison between $S_z$ and $|H|^2$ for the geometry described in Fig.11. We have used non-dimensional units, i.e. we take $\varepsilon_0=\mu_0=c=1$ and unitary amplitude for the electric and magnetic field of the plane, monochromatic, TM polarized wave incident on the screen. In those non-dimensional units the intensity carried by the plane wave is $S_{z,pw}=0.5$.

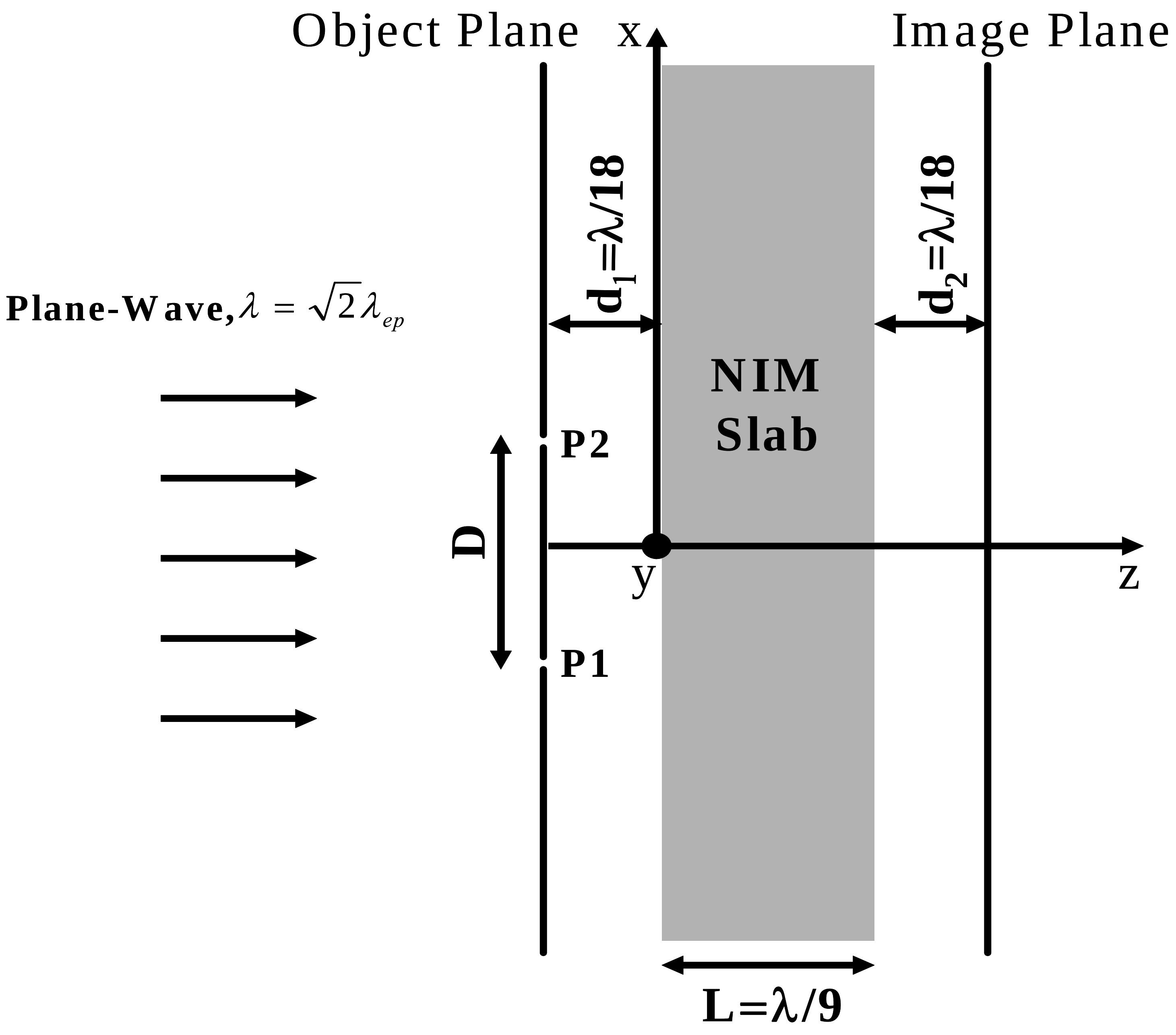

**Fig.1**

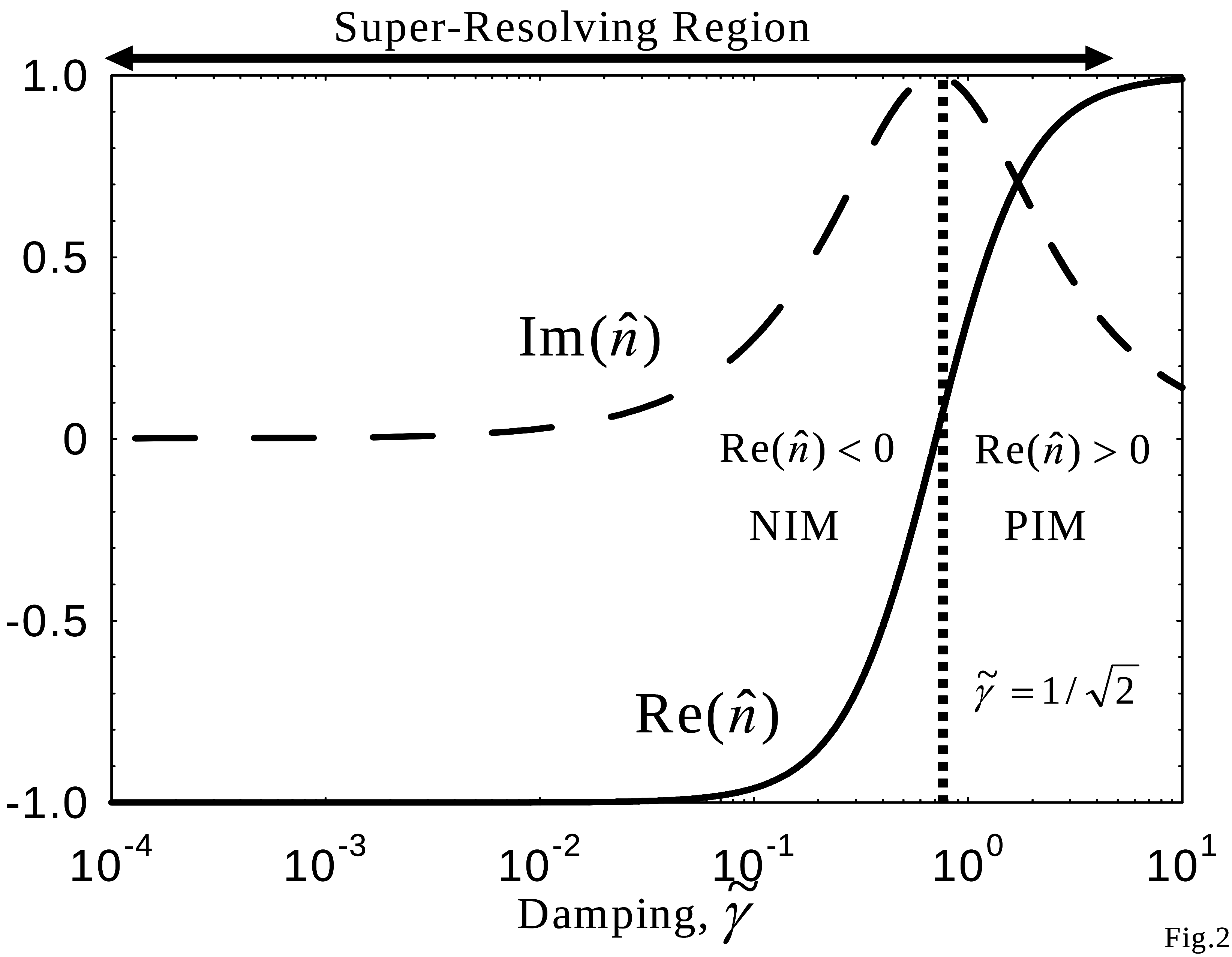

Fig.2(a)

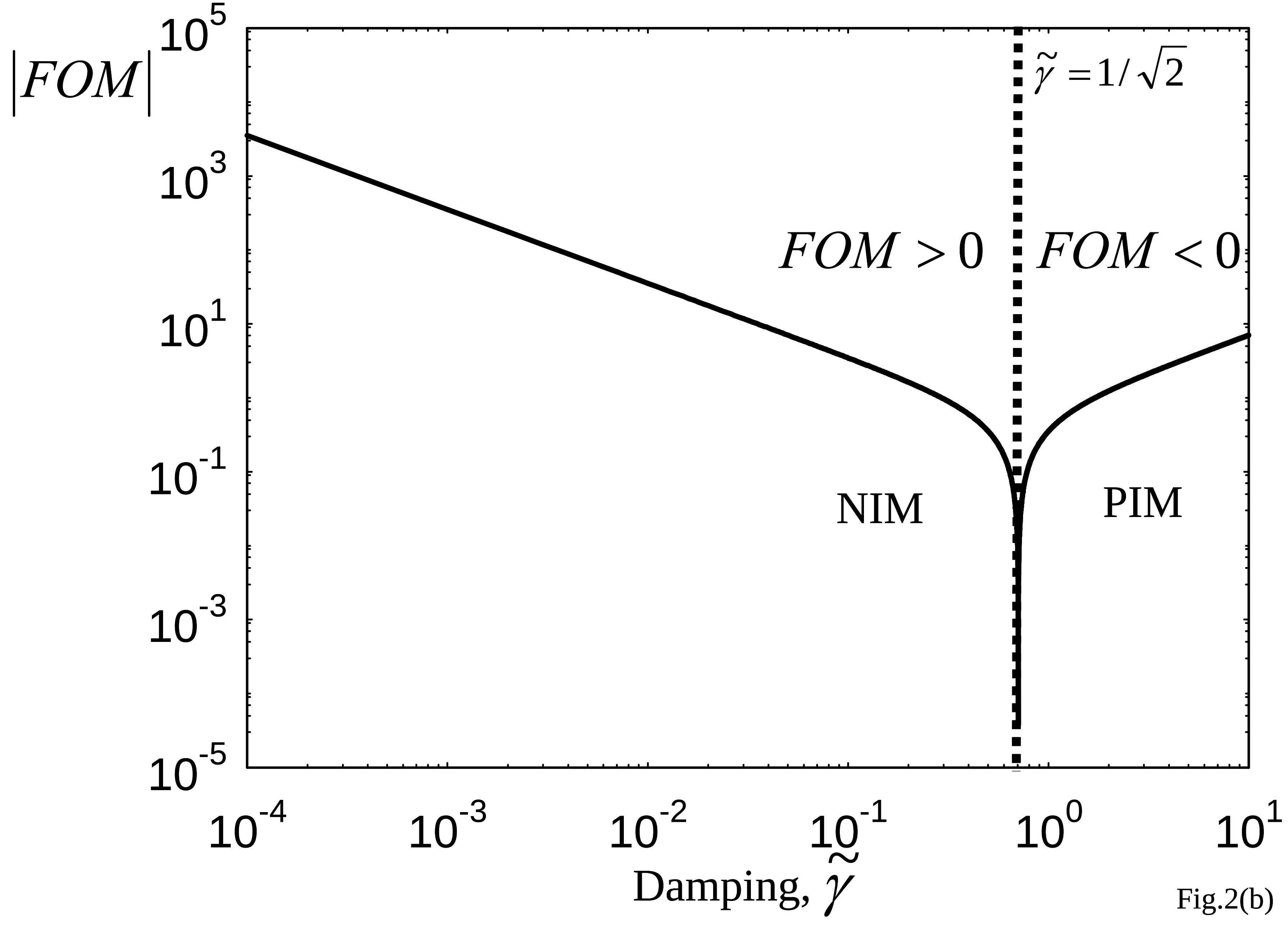

Fig.2(b)

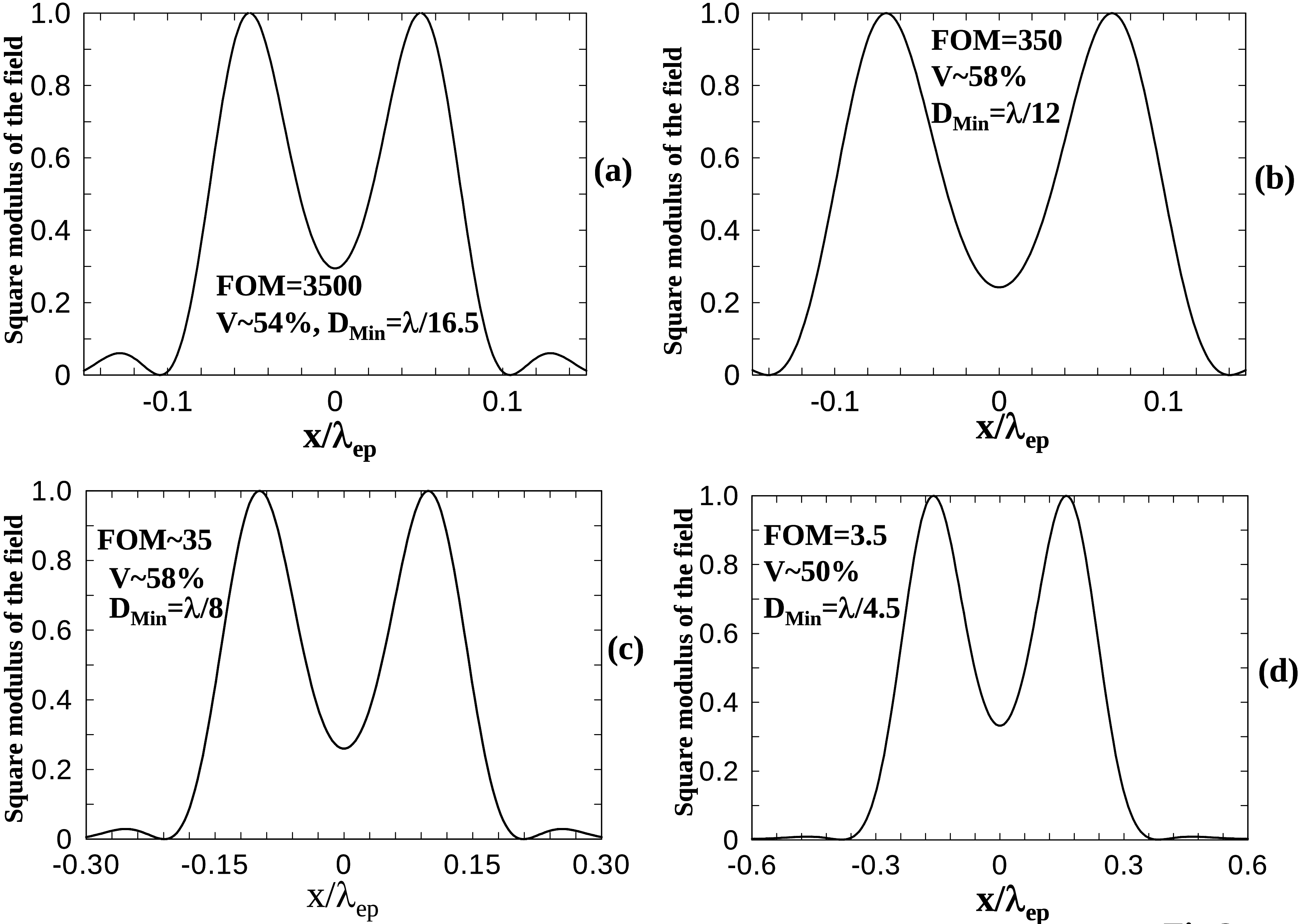

Fig.3

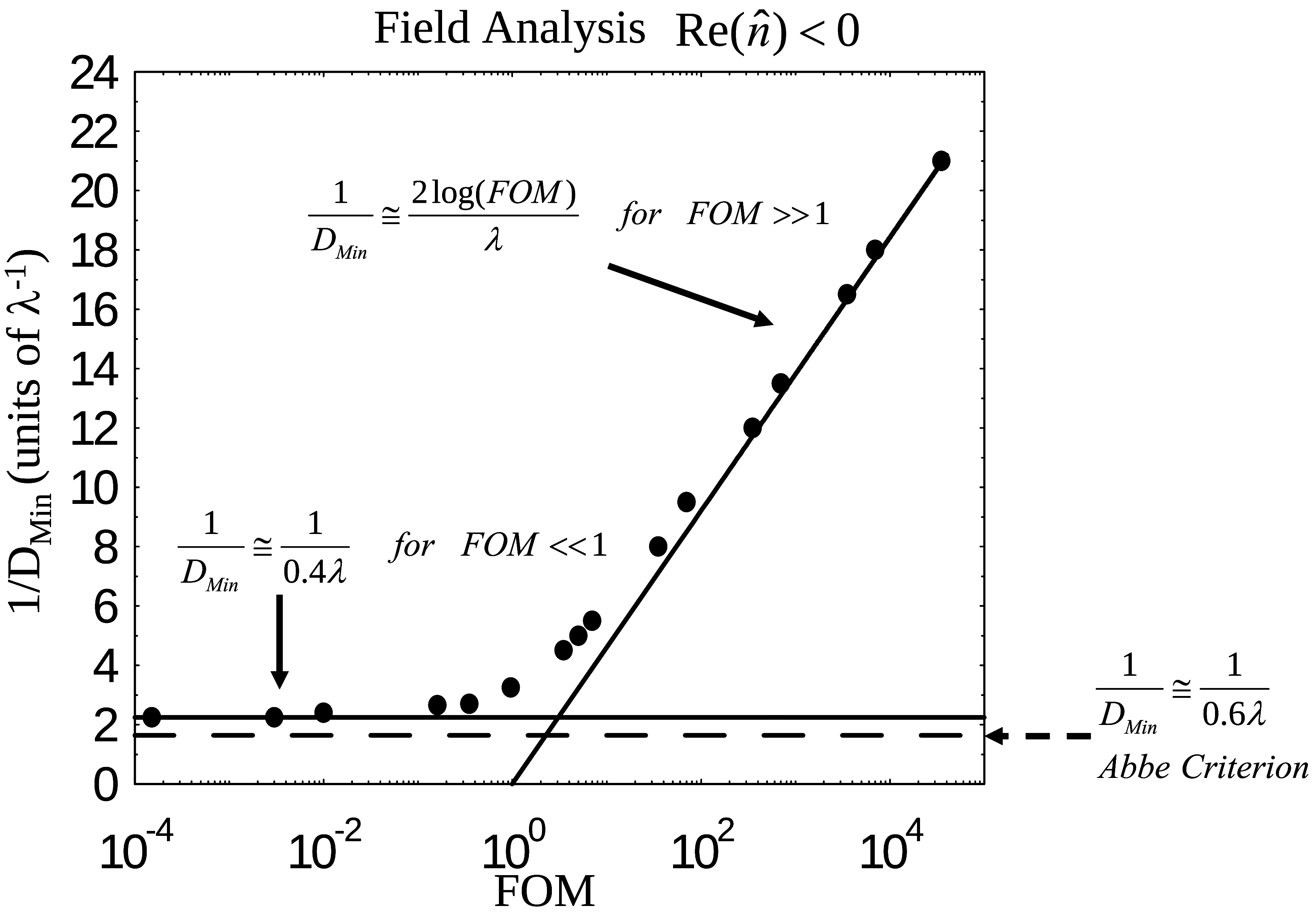

Field Analysis $\mathrm{Re}(\hat{n}) < 0$
$\frac{1}{D_{Min}} \cong \frac{2\log(FOM)}{\lambda}$ for FOM >> 1
$\frac{1}{D_{Min}} \cong \frac{1}{0.4\lambda}$ for FOM << 1
$\frac{1}{D_{Min}} \cong \frac{1}{0.6\lambda}$
Abbe Criterion
$1/D_{Min}$ (units of $\lambda^{-1}$)
0
2
4
6
8
10
12
14
16
18
20
22
24
$10^{-4}$
$10^{-2}$
$10^{0}$
$10^{2}$
$10^{4}$
FOM

**Fig.4**

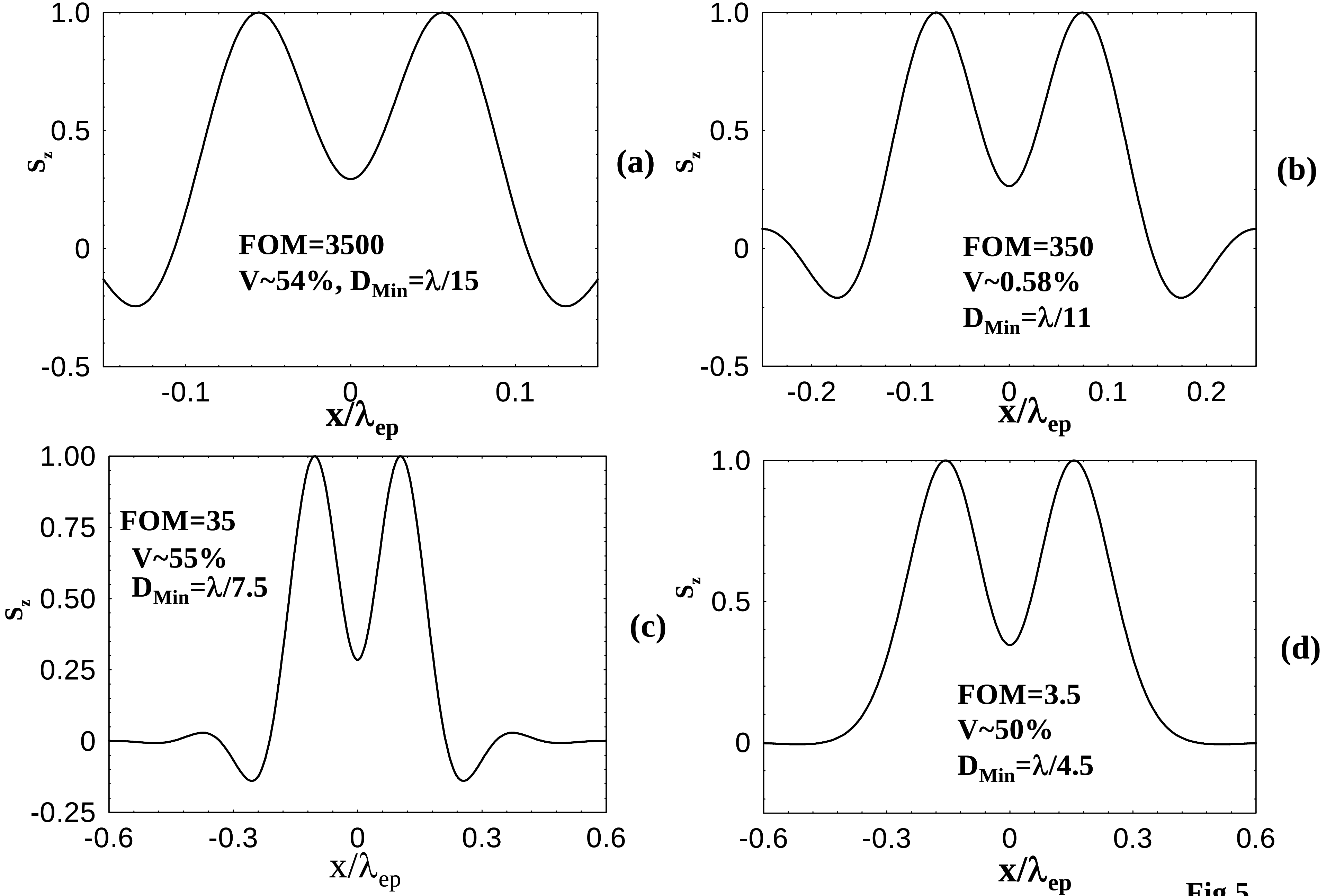

Fig.5

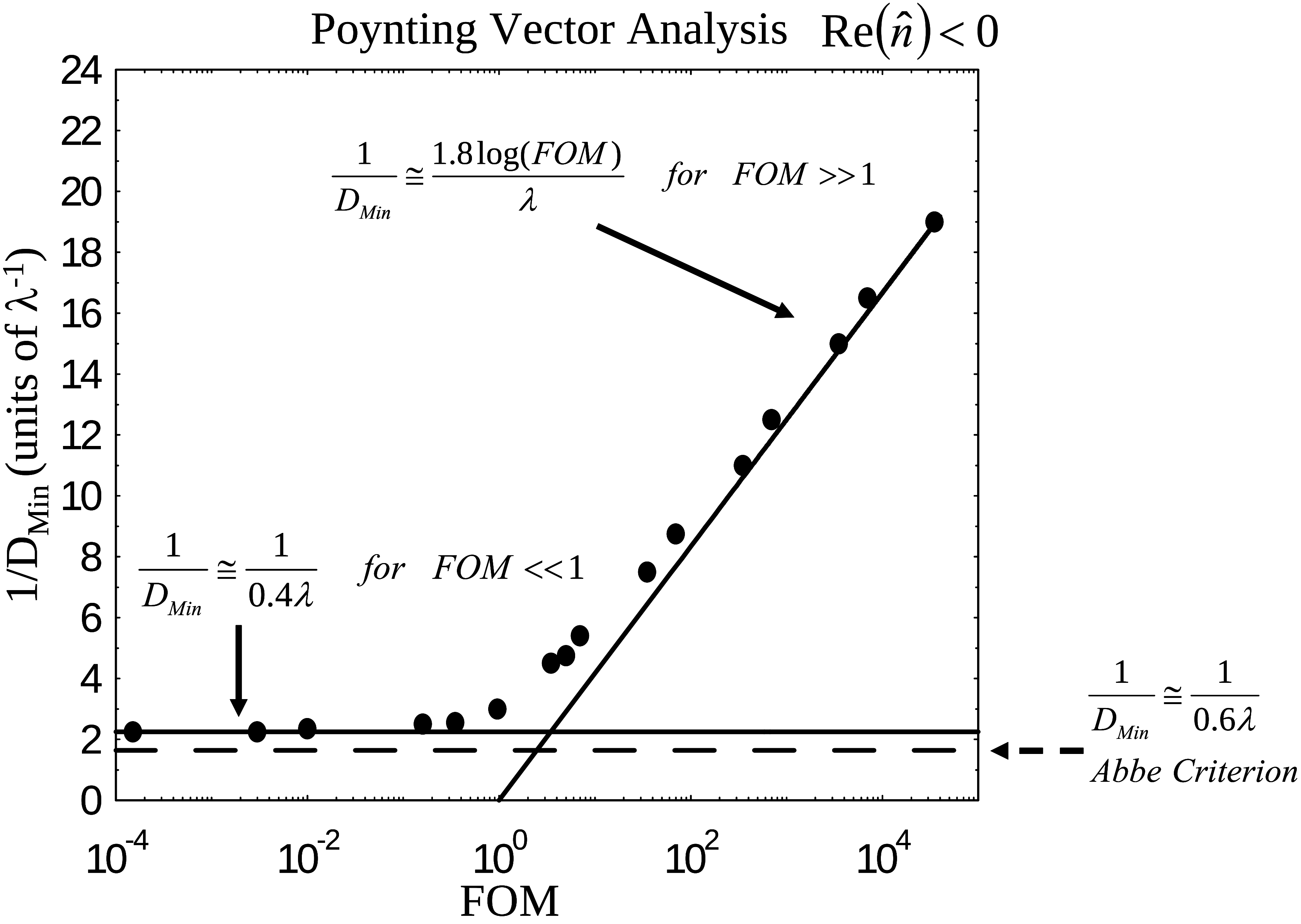

Poynting Vector Analysis $\mathrm{Re}(\hat{n}) < 0$
$\frac{1}{D_{Min}} \cong \frac{1.8\log(FOM)}{\lambda}$ for $FOM >> 1$
$\frac{1}{D_{Min}} \cong \frac{1}{0.4\lambda}$ for $FOM << 1$
$\frac{1}{D_{Min}} \cong \frac{1}{0.6\lambda}$
Abbe Criterion
$1/D_{Min}$ (units of $\lambda^{-1}$)
24
22
20
18
16
14
12
10
8
6
4
2
0
$10^{-4}$
$10^{-2}$
$10^{0}$
$10^{2}$
$10^{4}$
FOM

**Fig.6**

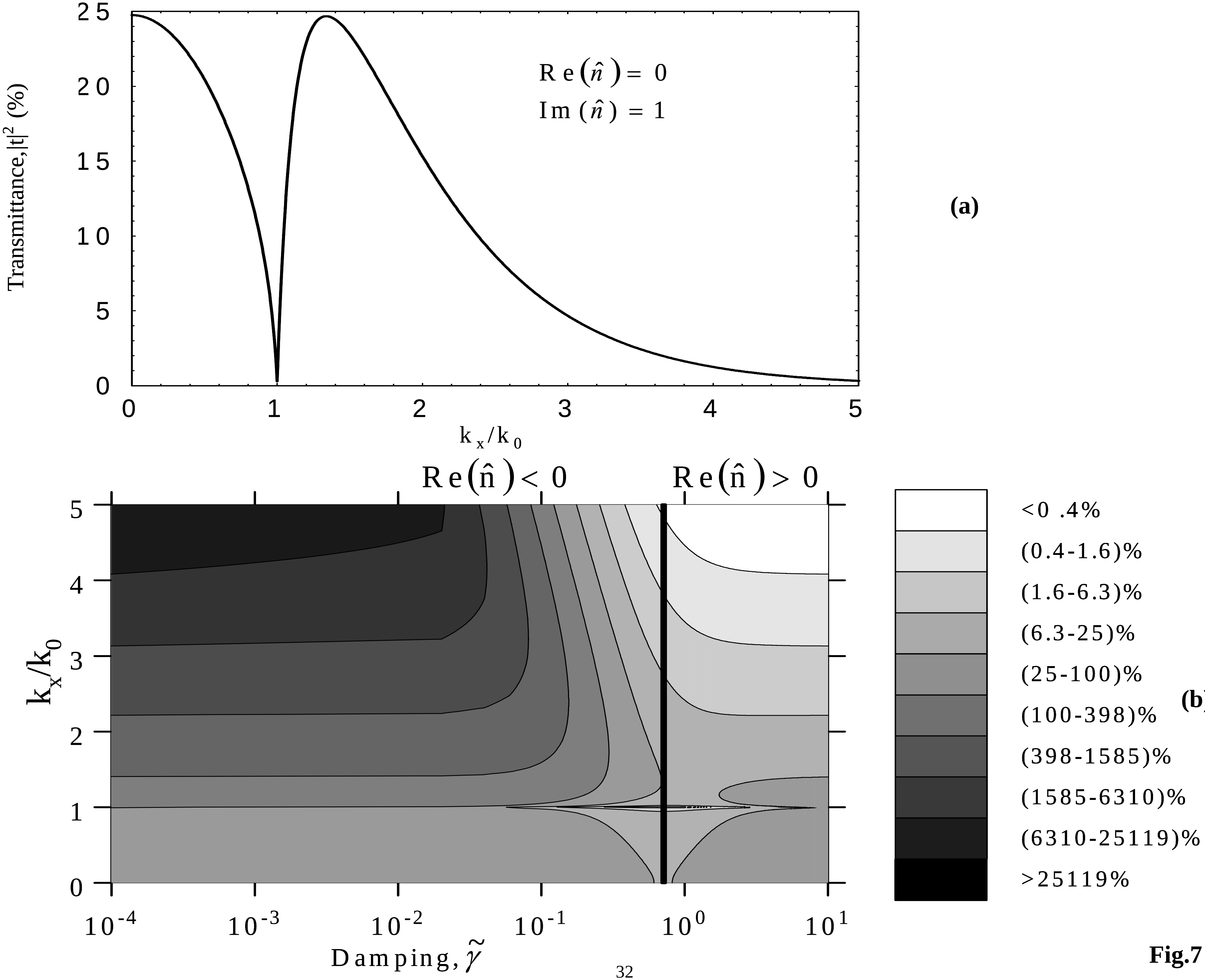

Fig.7

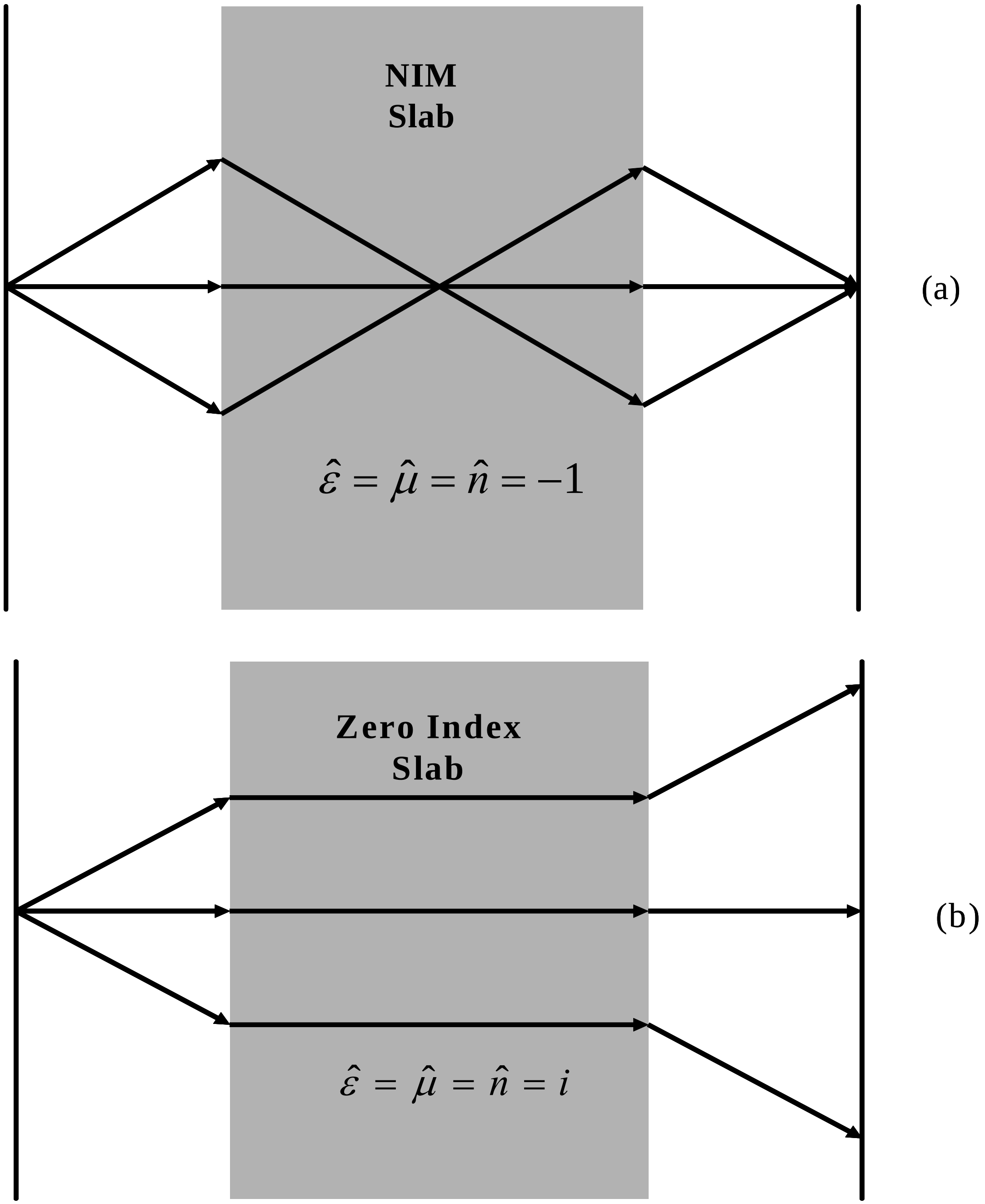
NIM
Slab
$\hat{\varepsilon} = \hat{\mu} = \hat{n} = -1$
(a)
Zero Index
Slab
$\hat{\varepsilon} = \hat{\mu} = \hat{n} = i$
(b)

**Fig.8**

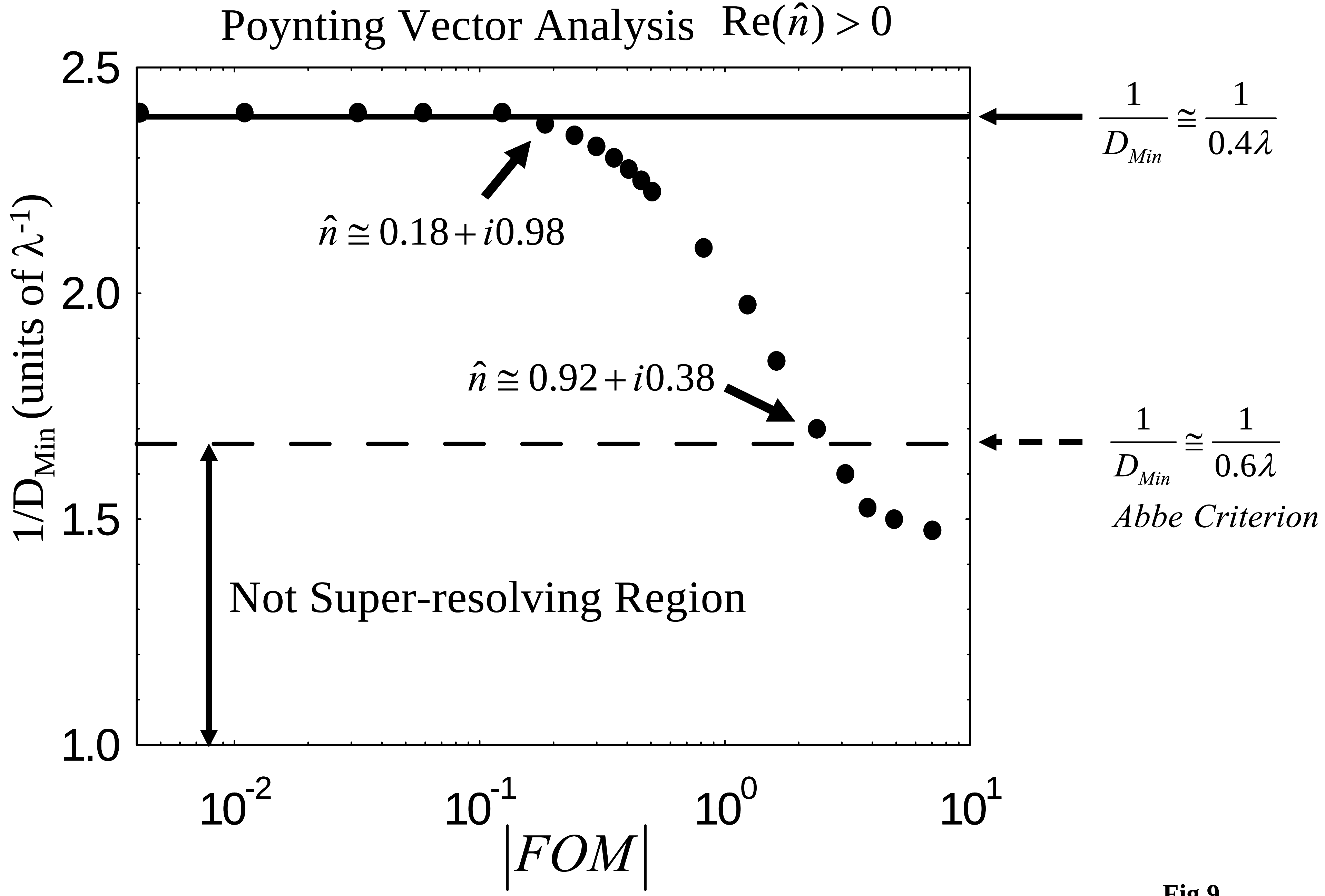

Poynting Vector Analysis $\mathrm{Re}(\hat{n}) > 0$
2.5
2.0
1.5
1.0
$1/D_{Min}$ (units of $\lambda^{-1}$)
$\hat{n} \cong 0.18 + i0.98$
$\hat{n} \cong 0.92 + i0.38$
$\frac{1}{D_{Min}} \cong \frac{1}{0.4\lambda}$
$\frac{1}{D_{Min}} \cong \frac{1}{0.6\lambda}$
*Abbe Criterion*
Not Super-resolving Region
$10^{-2}$
$10^{-1}$
$10^{0}$
$10^{1}$
$|FOM|$

**Fig.9**

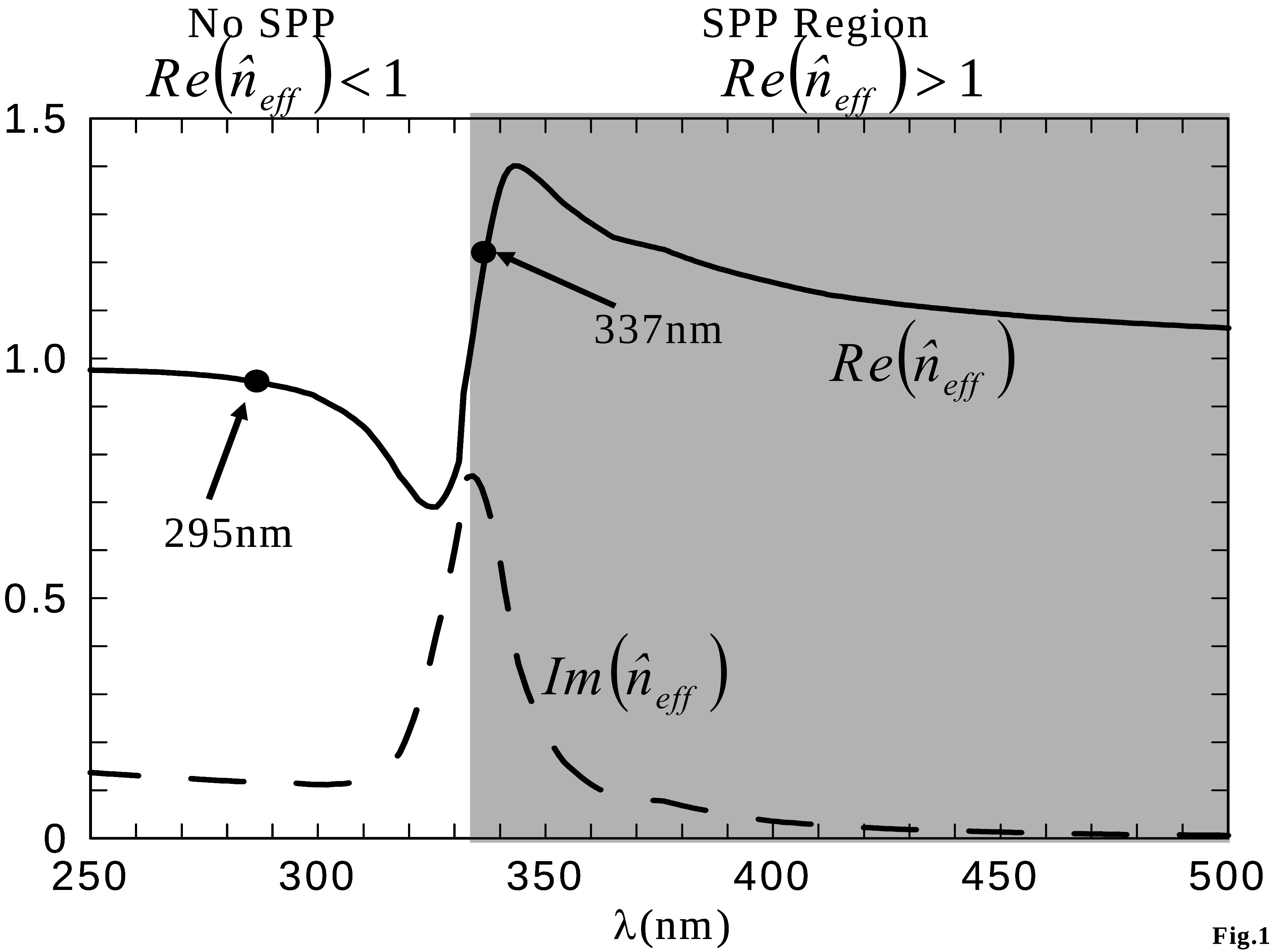

**Fig.10**

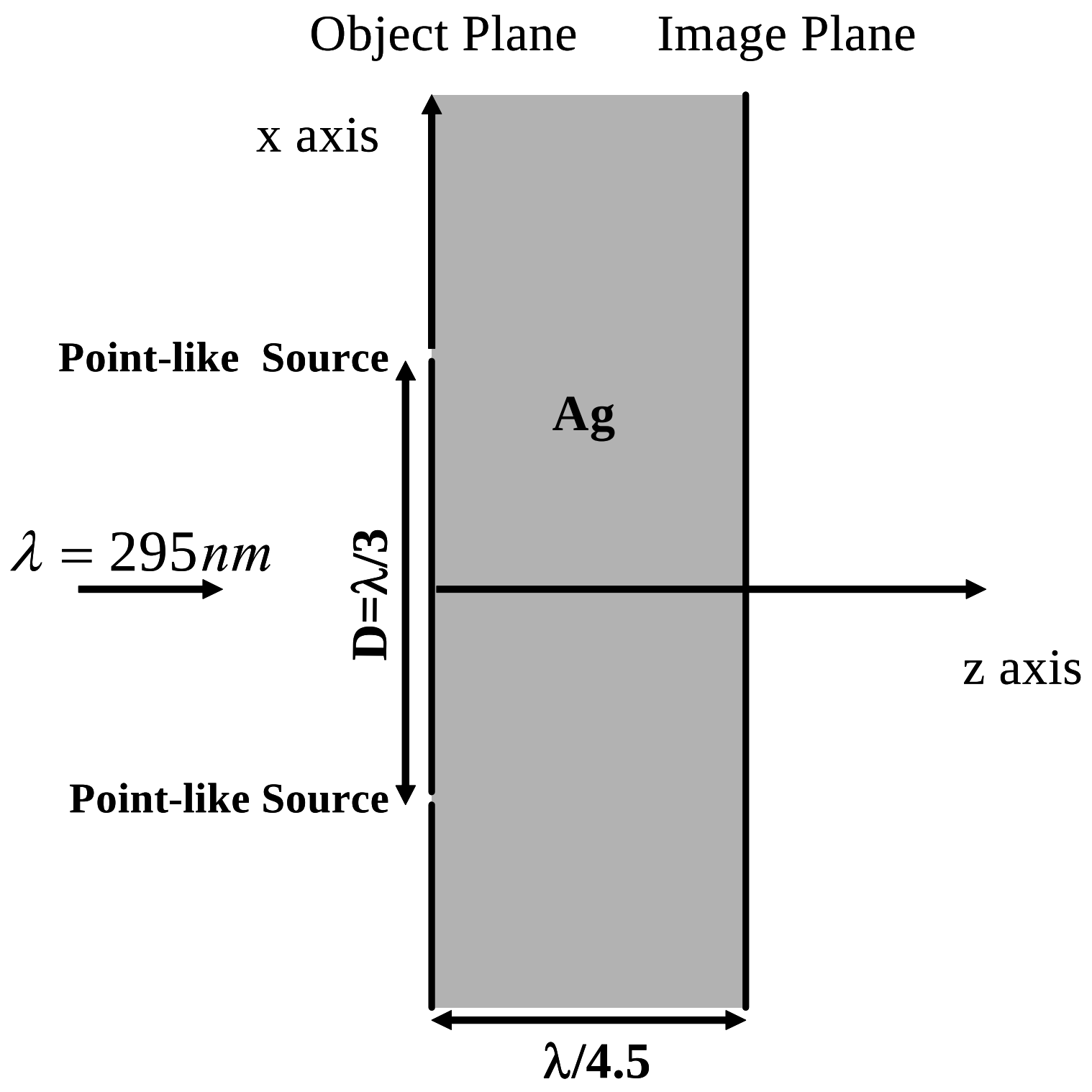

(a)

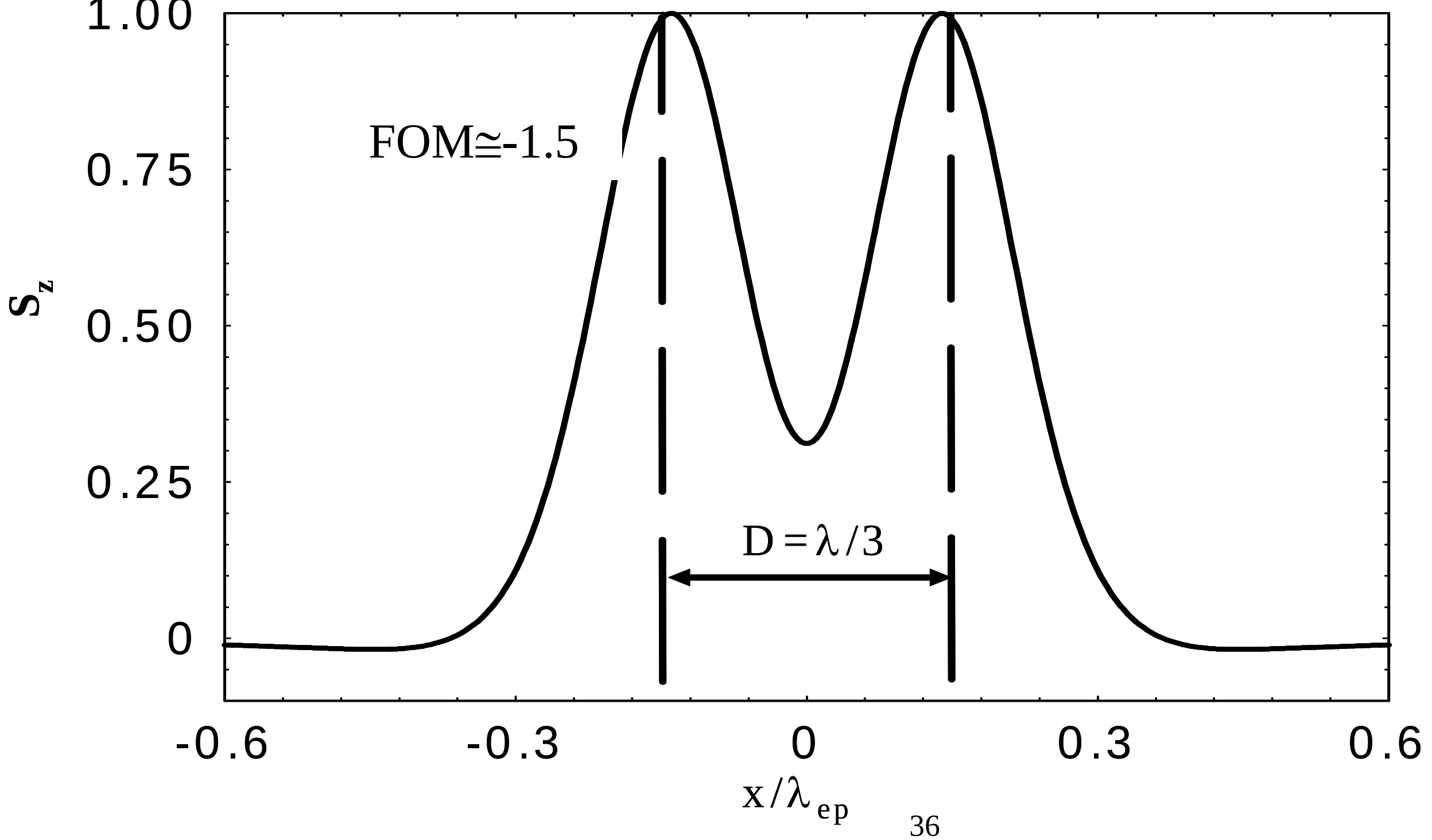

(b)

Fig.11

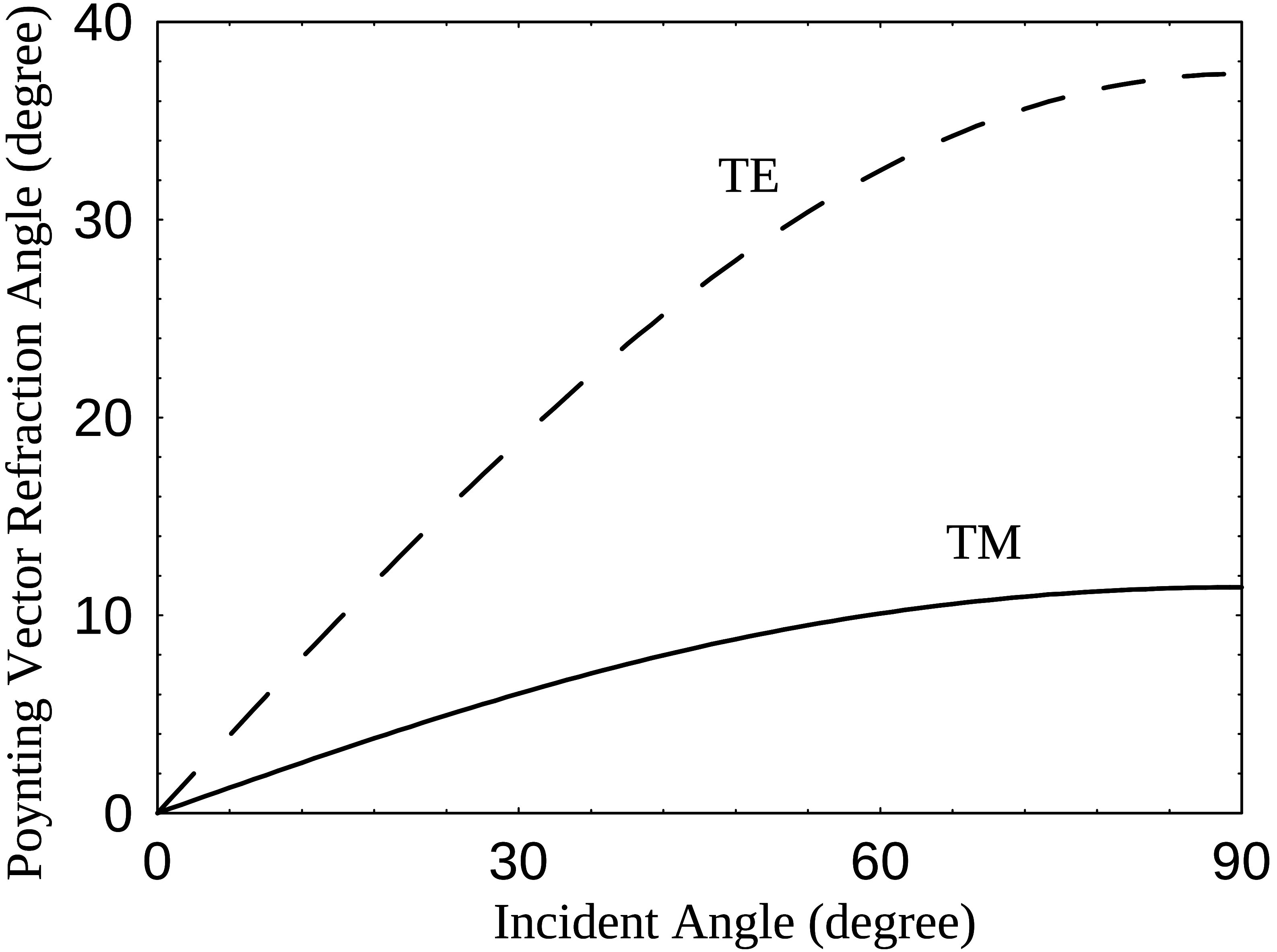

**Fig.12**

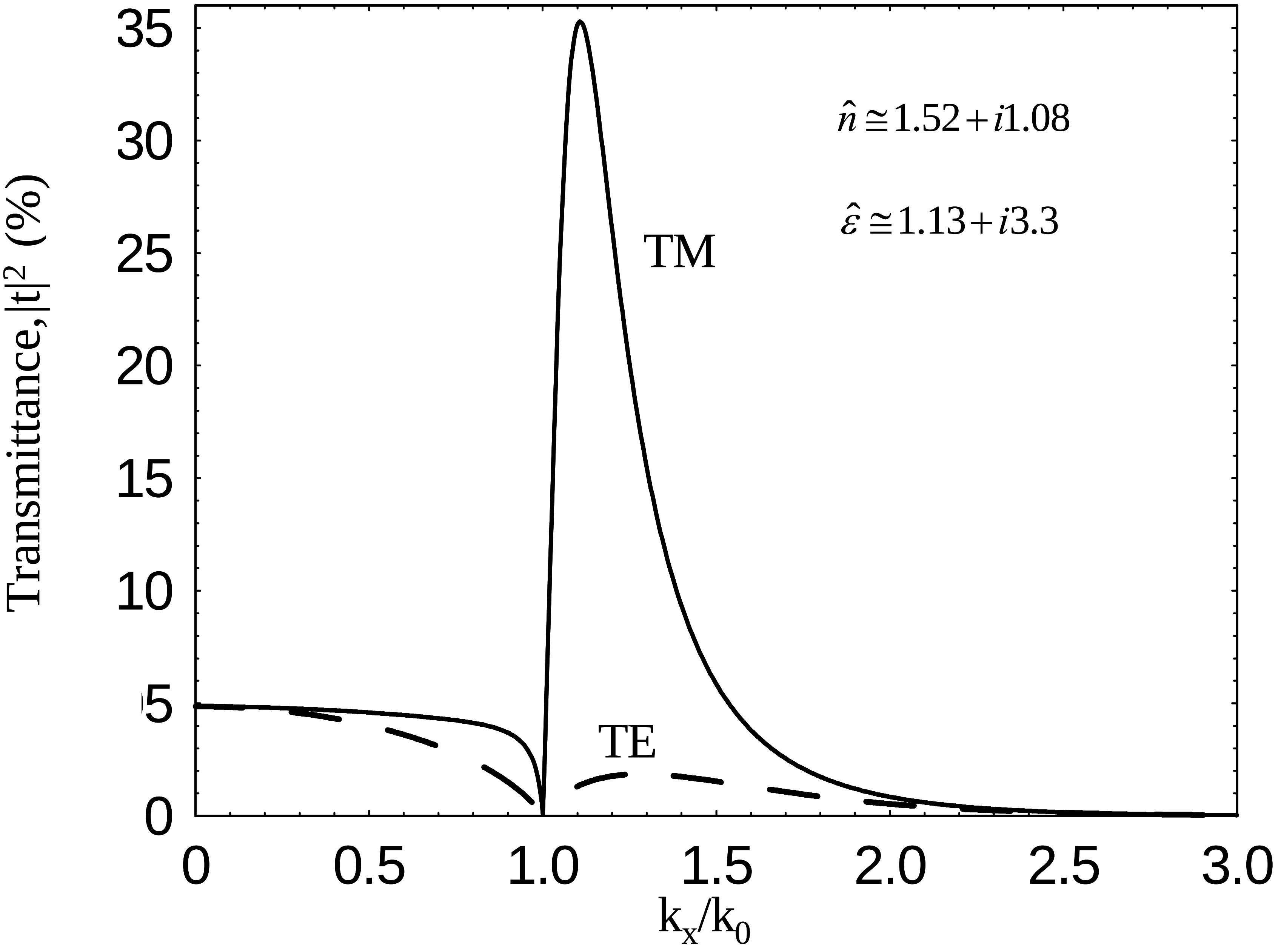

**Fig.13**

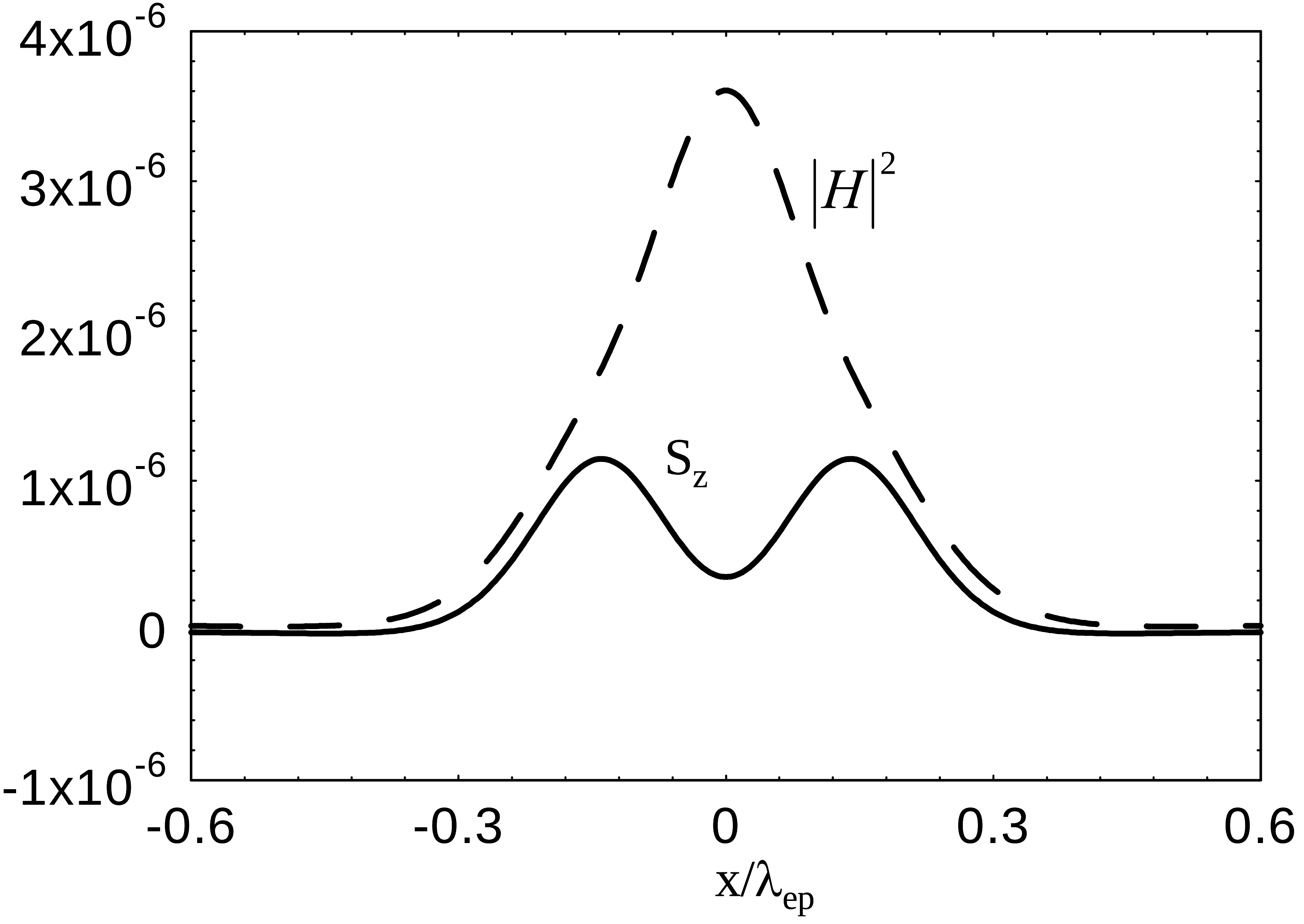

**Fig.14**